\documentclass[fleqn,usenatbib]{mnras} 
\usepackage{newtxtext,newtxmath}

\usepackage[T1]{fontenc}
\usepackage{ae,aecompl}
\usepackage{longtable}
\usepackage{pdflscape}
\usepackage{multirow}

\usepackage{graphicx}	
\usepackage{amsmath}	
\usepackage{xspace}
\usepackage{orcidlink}

\newcommand{\kms}{km\,s$^{-1}$\xspace}	
\newcommand{\Msun}{M$_\odot$\xspace}	

\title[Distance and age of Westerlund\,1 using Gaia-EDR3]{Distance and age of the massive stellar cluster Westerlund\,1. \\ I. Parallax method using Gaia-EDR3}

\author[F. Navarete]{
Felipe~Navarete$^{1,2}$\thanks{E-mail: felipe.navarete@noirlab.edu (FN)} \orcidlink{0000-0002-0284-0578},
Augusto~Damineli$^{2}$ \orcidlink{0000-0002-7978-2994},
{Aura~E.~Ramirez}$^{2}$ \orcidlink{0000-0002-6399-3498}, 
Danilo~F.~Rocha$^{3,4}$ \orcidlink{0000-0002-8225-7699},
\newauthor
Leonardo~A.~Almeida$^{4,5}$ \orcidlink{0000-0002-3817-6402}
\\
$^{1}$SOAR Telescope/NSF's NOIRLab, Avda Juan Cisternas 1500, 1700000, La Serena, Chile\\
$^{2}$Universidade de S\~ao Paulo, Instituto de Astronomia, Geof\'{\i}sica e Ci\^{e}ncias Atmosf\'{e}ricas, Rua do Mat\~ao, 1226, S\~ao Paulo - SP, 05508-090, Brazil \\
$^{3}$Observat\'orio Nacional, R. Gen. Jos\'e Cristino, 77 - Vasco da Gama, Rio de Janeiro - RJ, 20921-400, Brazil \\
$^{4}$Programa de P\'os-Gradua\c{c}\~ao em F\'isica, Universidade do Estado do Rio Grande do Norte, Mossor\'o - RN, 59610-210, Brazil \\
$^{5}$Escola de Ci\^encias e Tecnologia, Universidade Federal do Rio Grande do Norte, Natal - RN, 59072-970, Brazil }

\date{Accepted Aug 18th, 2022.}

\pubyear{2022}

\begin{document}
\label{firstpage}
\pagerange{\pageref{firstpage}--\pageref{lastpage}}
\maketitle

\begin{abstract}
{
Westerlund\,1 (Wd\,1) is one of the most massive young star clusters in the Milky Way. Although relevant for star formation and evolution, its fundamental parameters are not yet very well constrained. We aim to derive an accurate distance and provide constraints on the cluster age. We used the photometric and astrometric information available in the Gaia Early Data\,Release\,3 (Gaia-EDR3) to infer its distance of 4.06$^{+0.36}_{-0.34}$\,kpc.
Modelling of the eclipsing binary system W36, reported in Paper\,II, led to the distance of 4.03$\pm$0.25\,kpc, in agreement with the Gaia-EDR3 distance and, therefore, validating the parallax zero-point correction approach appropriate for red objects. The weighted average distance based on these two methods 
results in d$_{\rm wd1}$\,=\,4.05\,$\pm$\,0.20\,kpc ($m-M$\,=\,13.04$^{+0.11}_{-0.12}$\,mag), which has an unprecedented accuracy of 5\%.
Using the Binary Population and Spectral Synthesis (BPASS) models for the Red Supergiants with solar abundance, we derived an age of 10.7$\pm$1\,Myr, in excellent agreement with recent work by Beasor \& Davies (10.4$^{+1.3}_{-1.2}$\,Myr) based on MIST evolutionary models.
In Paper\,II, the age of W36B was reported to be 6.4$\pm$0.5\,Myr, supporting recent claims of a temporal spread of several Myrs for the star-forming process within Wd\,1 instead of a single monolithic starburst episode scenario.
}
\end{abstract}



\section{Introduction}
The formation and evolution of massive stars is poorly known because massive stellar clusters are extremely rare within distances that their stellar population can be resolved.  In this context, Westerlund\,1 (henceforth Wd\,1) is relevant, since it is potentially the most massive ($\sim$\,6\,$\times$\,$10^4$\,\Msun, \citealt{Portegies10} and references therein) young stellar cluster (4-11\,Myr, \citealt{beasor21,Brandner08,Gennaro11}) in our Galaxy. It is a local sibling of 30\,Dor, although a little denser and less massive than this one. Wd\,1 is located at $\sim$\,10\,times closer distance (3-5\,kpc - \citealt{Andersen17,Kothes07,Davies19,beasor21}), making it an ideal template for investigating the massive stellar formation and evolution process in the nearby Universe, but still at a considerable large distance and extinction ($A_{\rm V}$\,$\sim$\,11.5, \citealt{Damineli16, Hosek18}) that precluded its fully characterisation until now.
This massive cluster harbours a large population of confirmed massive stars evolved off the Main Sequence (MS): four Red Supergiants (RSG), 24 Wolf-Rayets (WR), more than 100 Blue Giant and Supergiants OB stars, one magnetar, one Luminous Blue Variable (LBV), one B[e], and six Yellow Hypergiants (YHG) - a short-lived phase \citep{Clark05, Crowther06, Muno06, Negueruela10}.
Given its large population of massive stars, Wd\,1 is expected to generate more than a thousand supernovae explosions, producing a significant number of X- and $\gamma$-rays binaries, and the coalescence of some of them will likely generate gravitational waves \citep{Abbot20}.
Indeed, several X-ray binaries were already identified due to thermal X-rays produced by wind-wind collision in massive binaries \citep{Skinner06}.
The fraction of binaries is still not precisely known but is estimated to be a relatively large fraction of the Wd\,1 stellar population ($>$\,70\%, \citealt{Crowther06}, \citealt{Clark19};  $>$\,40\%, Ritchie et al. 2021 - submitted)). Binarity among massive stars is relevant since $>$\,70\% of them will exchange mass with the companion, frequently leading to a binary merger \citep{sana12} with potential for producing gravitational waves.

The question of whether Wd\,1 serves as a ``golden-standard'' laboratory for massive stellar evolution \citep{Clark14} is still under debate. There is a large age discrepancy between different methods for Wd\,1. The pre-MS sequence population indicates an age of $\sim$\,5-7\,Myr \citep{Andersen17}, much younger than that of the less-luminous RSG ($\sim$\,11\,Myr, \citealt{beasor21}, leaving the door open for more than a unique episode of star formation over time. Since extinction and distance are not well constrained for Wd\,1, absolute luminosities are uncertain, impacting the determination of individual stellar ages and, thus, the age of the cluster itself.

{The presence of RSGs and WRs in a coeval cluster would indicate a narrow range of possible ages in a scenario of stars evolving in isolation ($\sim$\,5\,Myr, \citealt{Clark05}). However, binary interactions shape the unique cohort of hypergiant stars, with mass-stripped primaries, rejuvenated secondaries, and merger products \citep{Clark20}. \citet{beasor21} proposed that stars in Wd\,1 formed over a period of several Myr, similar to R136/NGC\,2070, based on the evolved population \citep{Schneider18}. However, this cluster shows many sub-clusters distributed throughout its confines which does not seem to be the case for Wd\,1. The same authors point to the difference between the HRD of Wd\,1 and RSGC1 \citep{davies08}. Although RSGC1 has a total mass and the age of RSGs similar to those in Wd\,1, it does not show any WR and is prone to represent a single starburst.}

The reddening law and extinction were accurately determined by \citet[][hereafter D16]{Damineli16} using colour indices of bright cluster members calibrated from spectral types reported by \citet{Clark05,Ritchie09,Negueruela10, Clark20} and $JHK_{s}$ photometry measurements combining telescopes of different diameters (4-m, 60-cm and mask aperture of 2-cm). \citet{Hosek18} used $JHK_{s}$ photometry and modelled the colour excesses of MS members by adopting a 5\,Myr isochrone. They found good agreement between the average extinction to the cluster ($A_{\rm Ks}$\,=\,0.78$\pm$0.16\,mag) with that reported by D16 ($A_{\rm Ks}$\,=\,0.74\,$\pm$\,0.01\,mag). \citet{Hosek18} also derived a steep optical/NIR reddening law similar to the of D16 and showed that it is not compatible with previous ones, even with those tuned to the inner Galaxy (Galactic Bulge, outside of the Galactic Plane), such as reported by \citet{Nishiyama09}. To date, the reddening law and the mean $A_{\rm Ks}$ extinction are the best well-constrained parameters for the Wd\,1 cluster.

Previous distance determinations, however, have not converged to a $\sim$\,10\% accuracy, which is an uncertainty level desired for this case. 
The first suggestion for the Wd\,1 distance was made by \citet{Westerlund61} by assuming it was located on the Sagittarius arm, at d\,$>$\,1.4\,kpc. \citet{Kothes07} analysed the velocity of \ion{H}{i} absorption features on a Galactic rotation curve model to derive the Wd\,1 distance as d\,=\,3.9$\pm$0.7\,kpc. Other distances were derived through colour-magnitude diagram (CMD) analysis, using isochrone fitting procedures to evaluate the distance modulus for the cluster. \citet{Westerlund87} reported d\,=\,5.0$^{+0.5}_{-1.0}$\,kpc, while \citet{Piatti98} derived d\,=\,1.0$\pm$0.4\,kpc. \citet{Clark05} suggested 2\,$<$\,d\,$<$\,5.5\,kpc through calibrated luminosity of YHG and limiting luminosity for RSGs to be lower than the  Humphreys-Davidson limit. The major issue in \citet{Clark05} distance determinations is related to the assumption of a standard reddening law ($R_{\rm V}$\,=\,3.1, \citealt{Cardelli89}) towards the cluster, in contradiction with their photometry which indicated that OB stars had a non-standard reddening. The minimum distance (2\,kpc) was based on the non-detection of radio-emission from WR stars. \citet{Crowther06} used $JHK_s$ photometry of WR stars obtaining d\,$\sim$\,5.5\,kpc. Their adoption of \citet{Indebetouw05} reddening law is a weak point since it does not apply to inner regions of the Galactic plane (see D16).
\citet{Negueruela10} used OB supergiants and isochrone fitting to derive a distance of d\,$\approx$\,5\,kpc. Despite those results being based on accurate spectral classification of the sources, they have uncertainties from using the \citet{Rieke85} reddening law and the assumption those are single stars, while many OB supergiants are probably binaries with similar brightness companions \citep{Crowther06, Clark19}.
\citet{Gennaro11} reported d\,=\,4.0$\pm$0.2\,kpc based on fit to the isochrones of pre-MS stars. \citet{Hosek18} obtained d\,=\,4.4$\pm$0.3\,kpc by fitting the MS assuming an age of 5\,Myr. The weakness of the isochrone fitting method is the large degeneracy between extinction and distance, in addition to the strong dependency on the adopted reddening law.  

Astrometric studies opened very promising horizons with the Gaia mission. Indeed, astrometric measurements (position, proper motion and parallaxes) are independent of age, extinction and reddening law. %
Unfortunately, the Gaia performance for objects at distances of $\sim$kpc scales, especially in dusty regions in the Galactic Center direction, is still impacted because parallax zero-points are still uncertain for very red colour indices \citep{Lindegren21b}.
For Wd\,1, the systematic effects are also affected by crowding, individual motions within the cluster, and motions due to binary orbits.
To test the reliability of the second Data Release of the Gaia mission (Gaia\,DR2) \citet{Aghakhanloo20} used Bayesian inference to derive the distance to both the Wd\,1 cluster members and the surrounding Galactic field stars, obtaining d\,=\,2.6$^{+0.6}_{-0.4}$\,kpc for Wd\,1. \citet{Aghakhanloo21} performed a similar study using the Gaia Early Data Release 3 (Gaia-EDR3), obtaining d\,=\,2.8$^{+0.7}_{-0.6}$\,kpc. 
Based on the parallax of OB-type stars in the Gaia\,DR2, \citet{Davies19} reported d\,=\,3.87$^{+0.95}_{-0.64}$\,kpc.
\citet{Rate20} used a sample of 18 WRs and OB stars present in Gaia\,DR2 to derive the distance of d\,$\sim$\,3\,kpc. \citet{beasor21} used the same method as \citet{Davies19} using Gaia-EDR3 which shows an improved zero-point parallax determination \citep{Lindegren21b}, reporting the distance of d\,=\,4.12$^{+0.66}_{-0.33}$\,kpc. In order to maximise cluster membership, they selected OB-type stars within {5\arcmin} from the cluster center and excluded stars with proper motions larger than 10\,\kms, which is the approximate Virial velocity for a cluster with 10$^5$\,\Msun and size of 1\,pc.
{More recently, \citet{Negueruela22}, reported a distance of d\,=\,4.23$^{+0.23}_{-0.21}$\,kpc based on the astrometry and photometry of the massive stellar population of Wd\,1 from \citet{Clark20}.}
 
In this work, we fine-tune procedures to improve the accuracy of the distance determination to the Westerlund\,1 cluster using astrometry and photometry from the Gaia-EDR3 cluster members reported by \citet{Clark20}.
We also adopt the distance of the W36 binary system from \citet[][hereafter Paper\,II]{Rocha22}, $d_{\rm w36}$\,=\,4.03\,$\pm$\,0.25\,kpc, as an independent distance measurement to confirm the Gaia-EDR3 results.

This manuscript is organised as follows. 
In Sec.\,\ref{sect_data}, we present the Gaia-EDR3 data used for deriving the distance to the Wd\,1 cluster and the $JHK_s$ photometry of the four RSGs.
Section\,\ref{sect_results} presents the main results, including the analysis of the Gaia-EDR3 data and the distance to the Wd\,1 cluster, followed by the age determination based on the Red Supergiant stars.
We discuss the implications of our results in Section\,\ref{sect_discussion}, and we summarise our results in Sect.\,\ref{sect_summary}.
 
\section{Introduction}
The formation and evolution of massive stars is poorly known because massive stellar clusters are extremely rare within distances that their stellar population can be resolved.  In this context, Westerlund\,1 (henceforth Wd\,1) is relevant, since it is potentially the most massive ($\sim$\,6\,$\times$\,$10^4$\,\Msun, \citealt{Portegies10} and references therein) young stellar cluster (4-11\,Myr, \citealt{beasor21,Brandner08,Gennaro11}) in our Galaxy. It is a local sibling of 30\,Dor, although a little denser and less massive than this one. Wd\,1 is located at $\sim$\,10\,times closer distance (3-5\,kpc - \citealt{Andersen17,Kothes07,Davies19,beasor21}), making it an ideal template for investigating the massive stellar formation and evolution process in the nearby Universe, but still at a considerable large distance and extinction ($A_{\rm V}$\,$\sim$\,11.5, \citealt{Damineli16, Hosek18}) that precluded its fully characterisation until now.
This massive cluster harbours a large population of confirmed massive stars evolved off the Main Sequence (MS): four Red Supergiants (RSG), 24 Wolf-Rayets (WR), more than 100 Blue Giant and Supergiants OB stars, one magnetar, one Luminous Blue Variable (LBV), one B[e], and six Yellow Hypergiants (YHG) - a short-lived phase \citep{Clark05, Crowther06, Muno06, Negueruela10}.
Given its large population of massive stars, Wd\,1 is expected to generate more than a thousand supernovae explosions, producing a significant number of X- and $\gamma$-rays binaries, and the coalescence of some of them will likely generate gravitational waves \citep{Abbot20}.
Indeed, several X-ray binaries were already identified due to thermal X-rays produced by wind-wind collision in massive binaries \citep{Skinner06}.
The fraction of binaries is still not precisely known but is estimated to be a relatively large fraction of the Wd\,1 stellar population ($>$\,70\%, \citealt{Crowther06}, \citealt{Clark19};  $>$\,40\%, Ritchie et al. 2021 - submitted)). Binarity among massive stars is relevant since $>$\,70\% of them will exchange mass with the companion, frequently leading to a binary merger \citep{sana12} with potential for producing gravitational waves.

The question of whether Wd\,1 serves as a ``golden-standard'' laboratory for massive stellar evolution \citep{Clark14} is still under debate. There is a large age discrepancy between different methods for Wd\,1. The pre-MS sequence population indicates an age of $\sim$\,5-7\,Myr \citep{Andersen17}, much younger than that of the less-luminous RSG ($\sim$\,11\,Myr, \citealt{beasor21}, leaving the door open for more than a unique episode of star formation over time. Since extinction and distance are not well constrained for Wd\,1, absolute luminosities are uncertain, impacting the determination of individual stellar ages and, thus, the age of the cluster itself.

{The presence of RSGs and WRs in a coeval cluster would indicate a narrow range of possible ages in a scenario of stars evolving in isolation ($\sim$\,5\,Myr, \citealt{Clark05}). However, binary interactions shape the unique cohort of hypergiant stars, with mass-stripped primaries, rejuvenated secondaries, and merger products \citep{Clark20}. \citet{beasor21} proposed that stars in Wd\,1 formed over a period of several Myr, similar to R136/NGC\,2070, based on the evolved population \citep{Schneider18}. However, this cluster shows many sub-clusters distributed throughout its confines which does not seem to be the case for Wd\,1. The same authors point to the difference between the HRD of Wd\,1 and RSGC1 \citep{davies08}. Although RSGC1 has a total mass and the age of RSGs similar to those in Wd\,1, it does not show any WR and is prone to represent a single starburst.}

The reddening law and extinction were accurately determined by \citet[][hereafter D16]{Damineli16} using colour indices of bright cluster members calibrated from spectral types reported by \citet{Clark05,Ritchie09,Negueruela10, Clark20} and $JHK_{s}$ photometry measurements combining telescopes of different diameters (4-m, 60-cm and mask aperture of 2-cm). \citet{Hosek18} used $JHK_{s}$ photometry and modelled the colour excesses of MS members by adopting a 5\,Myr isochrone. They found good agreement between the average extinction to the cluster ($A_{\rm Ks}$\,=\,0.78$\pm$0.16\,mag) with that reported by D16 ($A_{\rm Ks}$\,=\,0.74\,$\pm$\,0.01\,mag). \citet{Hosek18} also derived a steep optical/NIR reddening law similar to the of D16 and showed that it is not compatible with previous ones, even with those tuned to the inner Galaxy (Galactic Bulge, outside of the Galactic Plane), such as reported by \citet{Nishiyama09}. To date, the reddening law and the mean $A_{\rm Ks}$ extinction are the best well-constrained parameters for the Wd\,1 cluster.

Previous distance determinations, however, have not converged to a $\sim$\,10\% accuracy, which is an uncertainty level desired for this case. 
The first suggestion for the Wd\,1 distance was made by \citet{Westerlund61} by assuming it was located on the Sagittarius arm, at d\,$>$\,1.4\,kpc. \citet{Kothes07} analysed the velocity of \ion{H}{i} absorption features on a Galactic rotation curve model to derive the Wd\,1 distance as d\,=\,3.9$\pm$0.7\,kpc. Other distances were derived through colour-magnitude diagram (CMD) analysis, using isochrone fitting procedures to evaluate the distance modulus for the cluster. \citet{Westerlund87} reported d\,=\,5.0$^{+0.5}_{-1.0}$\,kpc, while \citet{Piatti98} derived d\,=\,1.0$\pm$0.4\,kpc. \citet{Clark05} suggested 2\,$<$\,d\,$<$\,5.5\,kpc through calibrated luminosity of YHG and limiting luminosity for RSGs to be lower than the  Humphreys-Davidson limit. The major issue in \citet{Clark05} distance determinations is related to the assumption of a standard reddening law ($R_{\rm V}$\,=\,3.1, \citealt{Cardelli89}) towards the cluster, in contradiction with their photometry which indicated that OB stars had a non-standard reddening. The minimum distance (2\,kpc) was based on the non-detection of radio-emission from WR stars. \citet{Crowther06} used $JHK_s$ photometry of WR stars obtaining d\,$\sim$\,5.5\,kpc. Their adoption of \citet{Indebetouw05} reddening law is a weak point since it does not apply to inner regions of the Galactic plane (see D16).
\citet{Negueruela10} used OB supergiants and isochrone fitting to derive a distance of d\,$\approx$\,5\,kpc. Despite those results being based on accurate spectral classification of the sources, they have uncertainties from using the \citet{Rieke85} reddening law and the assumption those are single stars, while many OB supergiants are probably binaries with similar brightness companions \citep{Crowther06, Clark19}.
\citet{Gennaro11} reported d\,=\,4.0$\pm$0.2\,kpc based on fit to the isochrones of pre-MS stars. \citet{Hosek18} obtained d\,=\,4.4$\pm$0.3\,kpc by fitting the MS assuming an age of 5\,Myr. The weakness of the isochrone fitting method is the large degeneracy between extinction and distance, in addition to the strong dependency on the adopted reddening law.  

Astrometric studies opened very promising horizons with the Gaia mission. Indeed, astrometric measurements (position, proper motion and parallaxes) are independent of age, extinction and reddening law. %
Unfortunately, the Gaia performance for objects at distances of $\sim$kpc scales, especially in dusty regions in the Galactic Center direction, is still impacted because parallax zero-points are still uncertain for very red colour indices \citep{Lindegren21b}.
For Wd\,1, the systematic effects are also affected by crowding, individual motions within the cluster, and motions due to binary orbits.
To test the reliability of the second Data Release of the Gaia mission (Gaia\,DR2) \citet{Aghakhanloo20} used Bayesian inference to derive the distance to both the Wd\,1 cluster members and the surrounding Galactic field stars, obtaining d\,=\,2.6$^{+0.6}_{-0.4}$\,kpc for Wd\,1. \citet{Aghakhanloo21} performed a similar study using the Gaia Early Data Release 3 (Gaia-EDR3), obtaining d\,=\,2.8$^{+0.7}_{-0.6}$\,kpc. 
Based on the parallax of OB-type stars in the Gaia\,DR2, \citet{Davies19} reported d\,=\,3.87$^{+0.95}_{-0.64}$\,kpc.
\citet{Rate20} used a sample of 18 WRs and OB stars present in Gaia\,DR2 to derive the distance of d\,$\sim$\,3\,kpc. \citet{beasor21} used the same method as \citet{Davies19} using Gaia-EDR3 which shows an improved zero-point parallax determination \citep{Lindegren21b}, reporting the distance of d\,=\,4.12$^{+0.66}_{-0.33}$\,kpc. In order to maximise cluster membership, they selected OB-type stars within {5\arcmin} from the cluster center and excluded stars with proper motions larger than 10\,\kms, which is the approximate Virial velocity for a cluster with 10$^5$\,\Msun and size of 1\,pc.
{More recently, \citet{Negueruela22}, reported a distance of d\,=\,4.23$^{+0.23}_{-0.21}$\,kpc based on the astrometry and photometry of the massive stellar population of Wd\,1 from \citet{Clark20}.}
 
In this work, we fine-tune procedures to improve the accuracy of the distance determination to the Westerlund\,1 cluster using astrometry and photometry from the Gaia-EDR3 cluster members reported by \citet{Clark20}.
We also adopt the distance of the W36 binary system from \citet[][hereafter Paper\,II]{Rocha22}, $d_{\rm w36}$\,=\,4.03\,$\pm$\,0.25\,kpc, as an independent distance measurement to confirm the Gaia-EDR3 results.

This manuscript is organised as follows. 
In Sec.\,\ref{sect_data}, we present the Gaia-EDR3 data used for deriving the distance to the Wd\,1 cluster and the $JHK_s$ photometry of the four RSGs.
Section\,\ref{sect_results} presents the main results, including the analysis of the Gaia-EDR3 data and the distance to the Wd\,1 cluster, followed by the age determination based on the Red Supergiant stars.
We discuss the implications of our results in Section\,\ref{sect_discussion}, and we summarise our results in Sect.\,\ref{sect_summary}.
 
\section{Data} \label{sect_data}

\subsection{Gaia-EDR3 archive}

We downloaded the list containing all sources of the Gaia-EDR3 catalogue within a {15\arcmin} radius around the central position of the Wd\,1 cluster (RA\,=\,16:47:04.00, Decl.\,=\,$-$45:51:04.9, \citealt{Clark05}), and a total of 36,116 sources were found. We kept only sources associated with {a Renormalised Unit Weight Error (RUWE) value $\leq$\,1.4 (for more details on the definition of RUWE, see \citealt{Lindegren18})}, and those detected above 3-$\sigma$ in all three photometric bands (BP, G and RP), leading to a final list of 25,501 objects. 

The new astrometric solution for Gaia-EDR3 \citep{Lindegren21,Lindegren21b}) divides the sources into three main groups:
$a)$ ``2-p'': objects with two parameters solved (i.e. position);
$b)$ ``5-p'': five parameters solved (i.e. position, parallax, proper motion); and
$c)$ ``6-p'': six parameters solved, including the source colour (effective wavelength) defined as \texttt{pseudocolour} in the Gaia catalogue.
The 6-p category encompasses the faintest sources, and/or objects with no reliable colour listed in Gaia\,DR2 \citep{Fabricius21}. Among the astrometric groups, the 5-p sources exhibit the most accurate astrometric solutions and, therefore, correspond to the best class of objects to investigate regions associated with high-extinction and/or at relatively large distances.

Among the selected 25,501 Gaia-EDR3 objects, 7,736 and 17,765 are associated with 5-p and 6-p astrometric solutions, respectively.

\subsubsection{Wd\,1 members in the Gaia-EDR3 catalogue} \label{sect_members}

The identification of cluster members in the Gaia-EDR3 catalogue was performed in two steps. First, we identified each member in one of the few catalogues which have been cross-matched with GAIA-DR3, such as the Guide Star Catalog II \citet[GSC-II][]{Lasker08} in visible wavelengths, and the Two Micron All Sky Survey \citet[2MASS][]{Skrutskie06} in the near-infrared. Then, we cross-matched the GSC-II IDs with the \texttt{gsc23\_best\_neighbour} table of the Gaia-EDR3 catalogue, and the 2MASS IDs with the \texttt{tmass\_best\_neighbour} table.
The positions of 166 stellar members of Westerlund\,1 reported by \citet{Clark20} were cross-matched with the GSC-II and the 2MASS catalogues (assuming a maximum offset of 1\arcsec), resulting in a list of 85 sources with a 2MASS counterpart and 105 sources with a GSC counterpart.

Then, the lists were cross-matched with the Gaia-EDR3, leading to a final list with 106 Wd\,1 members associated with a Gaia-EDR3 counterpart, from which 61 corresponds to 5-p sources (listed in Table\,\ref{table_Wd1members_Gaia5p}), 29 sources are 6-p (Table\,\ref{table_Wd1members_Gaia6p}), and 16 have Gaia-EDR3 counterparts with no astrometric information.
These last 16 objects were merged with 60 other Wd\,1 members with no Gaia-EDR3 counterparts (Table\,\ref{table_Wd1members_noGaia}).

\setlength{\tabcolsep}{4pt}
\begin{table*}
\caption{List of known Westerlund 1 members with Gaia EDR3 counterparts associated with a 5-parameter astrometric solution (5-p). The full table is available online at the CDS.}
\label{table_Wd1members_Gaia5p_short}
\begin{tabular}{llcccccccccccc}
\hline
\multicolumn{1}{c}{Name} & 
\multicolumn{1}{c}{GSC} &  \multicolumn{1}{c}{2MASS} &  \multicolumn{1}{c}{Gaia EDR3} &  \multicolumn{1}{c}{RUWE} &  \multicolumn{1}{c}{$\pi$} &   \multicolumn{1}{c}{$\pi_{\rm ZP}$} &  \multicolumn{1}{c}{$\mu_\alpha$} &  \multicolumn{1}{c}{$\mu_\delta$} &  \multicolumn{1}{c}{$\nu_{\rm eff}$}  \\
\multicolumn{1}{c}{} & 
\multicolumn{1}{c}{designation} &  \multicolumn{1}{c}{designation} &  \multicolumn{1}{c}{designation} &  \multicolumn{1}{c}{} &  \multicolumn{1}{c}{(mas)} &   \multicolumn{1}{c}{(mas)} &  \multicolumn{1}{c}{(mas\,yr$^{-1}$)} &  \multicolumn{1}{c}{(mas\,yr$^{-1}$)} &  \multicolumn{1}{c}{(\micron$^{-1}$)}  \\
\hline
W2a	&	--	&	16465971-4550513	&	5940106758703247360	&	0.94	&	0.264\,$\pm$\,0.044	&	$-$0.068	&	$-$2.015\,$\pm$\,0.053	&	$-$3.414\,$\pm$\,0.045	&	1.156		\\
W4	&	S8UV052685	&	16470142-4550373	&	5940106763014985088	&	0.96	&	0.180\,$\pm$\,0.059	&	$-$0.045	&	$-$2.237\,$\pm$\,0.073	&	$-$3.415\,$\pm$\,0.063	&	1.136 \\
W5	&	2MIU38SH	&	16470298-4550199	&	5940106797374726784	&	1.09	&	0.154\,$\pm$\,0.056	&	$-$0.073	&	$-$2.495\,$\pm$\,0.068	&	$-$3.467\,$\pm$\,0.059	&	1.151	\\
W6a	&	2MIU38SN	&	16470303-4550235	&	5940106797374726144	&	1.01	&	0.249\,$\pm$\,0.059	&	$-$0.076	&	$-$2.222\,$\pm$\,0.071	&	$-$3.803\,$\pm$\,0.062	&	1.150	\\
W7	&	S8UV052690	&	16470363-4550144	&	5940106793062992128	&	0.89	&	0.122\,$\pm$\,0.059	&	$-$0.014	&	$-$2.379\,$\pm$\,0.074	&	$-$3.713\,$\pm$\,0.063	&	1.131	\\
W8a	&	S8UV052686	&	16470480-4550251	&	5940106041460479488	&	0.88	&	0.119\,$\pm$\,0.061	&	$-$0.018	&	$-$2.376\,$\pm$\,0.075	&	$-$3.786\,$\pm$\,0.065	&	1.129	\\
W10	&	S8UV052706	&	16470334-4550346	&	5940106797366404480	&	1.00	&	0.137\,$\pm$\,0.063	&	$-$0.075	&	$-$1.989\,$\pm$\,0.075	&	$-$3.421\,$\pm$\,0.065	&	1.133	\\
W11	&	S8UV052697	&	--	&	5940106758703254656	&	0.95	&	0.144\,$\pm$\,0.052	&	$-$0.071	&	$-$2.480\,$\pm$\,0.064	&	$-$4.007\,$\pm$\,0.056	&	1.143		\\
W12a	&	S8UV052692	&	16470222-4550590	&	5940106763006638848	&	1.00	&	0.154\,$\pm$\,0.073	&	$-$0.006	&	$-$2.493\,$\pm$\,0.088	&	$-$3.692\,$\pm$\,0.076	&	1.114		\\
W13	&	S8UV052698	&	16470646-4550261	&	5940106037157928064	&	0.94	&	0.094\,$\pm$\,0.052	&	$-$0.071	&	$-$2.497\,$\pm$\,0.069	&	$-$4.112\,$\pm$\,0.060	&	1.147	\\
\hline
\end{tabular} \\
{\textbf{Notes:} The columns are as follows:
(1) Name of the source (the coordinates are listed in \citet{Clark19}); 
(2) designation of the source in the GSC 2.3 catalogue;
(3) designation of the source in the 2MASS PSC catalogue \citep{Skrutskie06};
(4) Identification of the Gaia EDR3 source;
(5) Renormalized Unit Weight Error;
(6) Parallax listed from the Gaia EDR3 and its error;
(7) Parallax zero-point correction derived from \citet{Lindegren21b};
(8)-(9) Proper motion in RA and Decl axis and their error;
(10) $\nu_{\rm eff}$.
}
\end{table*}

\setlength{\tabcolsep}{6pt} 

\subsubsection{Parallax zero-point correction: the analysis for Wd\,1 members} \label{sect_zp}

We corrected Gaia-EDR3 parallaxes by using the methodology described in \citet{Lindegren21b} and implemented in Python\footnote{\url{https://www.cosmos.esa.int/web/gaia/edr3-code}}, which estimates the parallax zero-point ($\pi_{\rm ZP}$)  separately for 5- and 6-parameter astrometric solutions as a function of the ecliptic latitude (\texttt{ecl\_lat}), the $G$-band magnitude (\texttt{phot\_g\_mean\_mag}), and the colour (effective wavenumber) for each source.
The correction is well-determined for objects exhibiting $G$-band magnitudes between 6 and 21\,mag, and effective wavenumber values between 1.1 and 1.9 for 5-p solutions (\texttt{nu\_eff\_used\_in\_astrometry}), and between 1.24 and 1.72 for 6-p solutions (\texttt{pseudocolour}).

\begin{figure}
    \centering
    {\includegraphics[width=\linewidth]{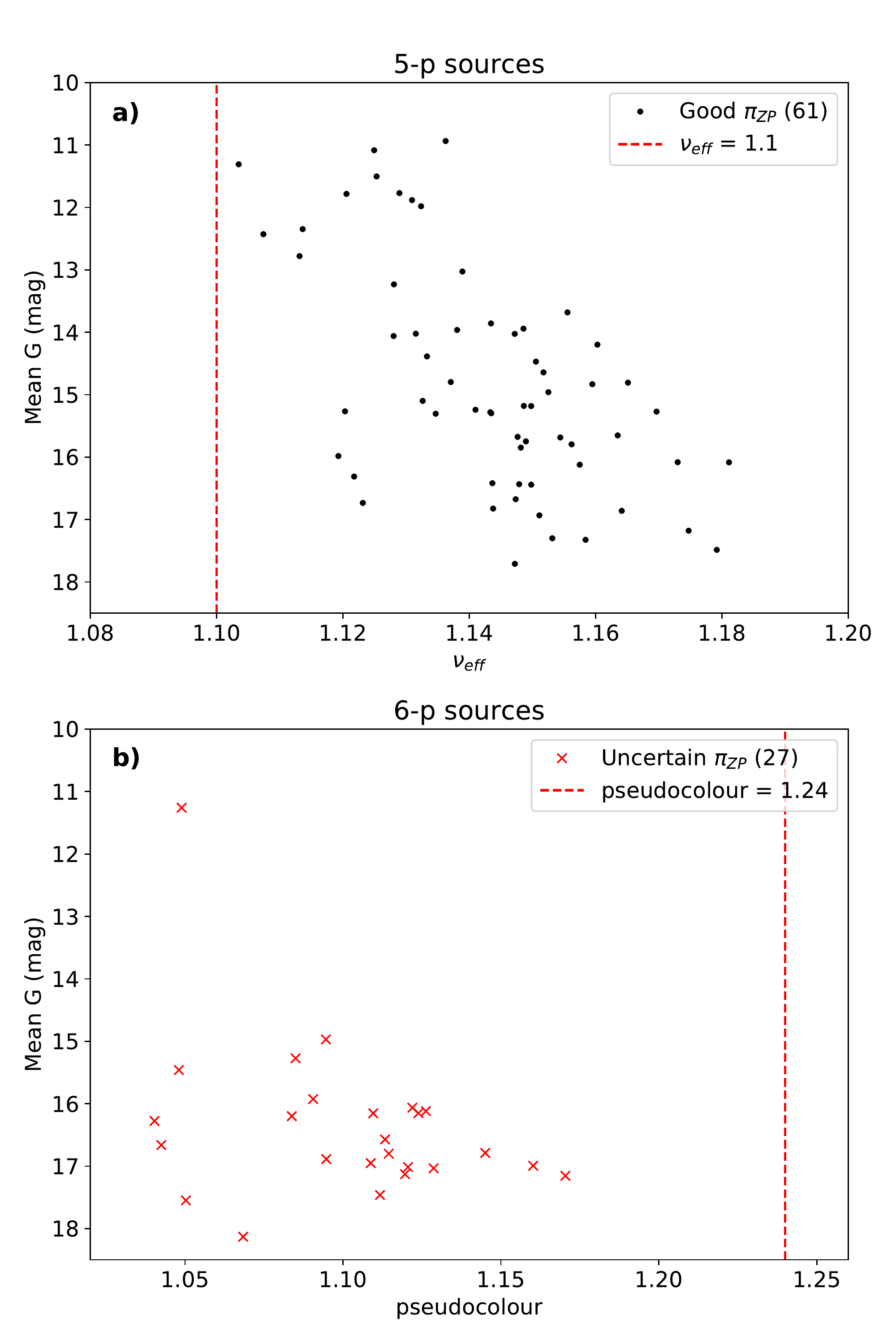}}
    \\[-1.0ex]
    \caption{Colour versus G-band mean magnitude for the known Wd\,1 members with Gaia-EDR3 counterparts exhibiting five (5-p, panel a) and six astrometric parameters solved (6-p, panel b). Black points indicate the objects with a reliable parallax zero-point correction, while red $\times$ symbols indicate sources located outside the limits for the correction. The vertical dashed lines are placed at $\nu_{\rm eff}$\,=\,1.1 (a) and pseudocolour\,=\,1.24 (b).}
    \label{fig_zeropoint_conditions}
\end{figure}

Figure\,\ref{fig_zeropoint_conditions} presents the distribution of the $G$-band magnitude against the effective wavenumber, for the known Wd\,1 members classified as 5-p (61 sources, panel a) and as 6-p objects (27, panel b).
In Fig.\,\ref{fig_zeropoint_conditions}\,(a), the distribution of the 5-p sources shows that all points are located within the limits for the proper determination of their $\pi_{\rm ZP}$ values. On the other hand, Fig.\,\ref{fig_zeropoint_conditions}\,(b) shows that all the 6-p sources exhibit pseudocolour values outside the limits for obtaining a proper ZP correction, leading to dubious values. 

\begin{figure}
    \centering
    {\includegraphics[width=\linewidth]{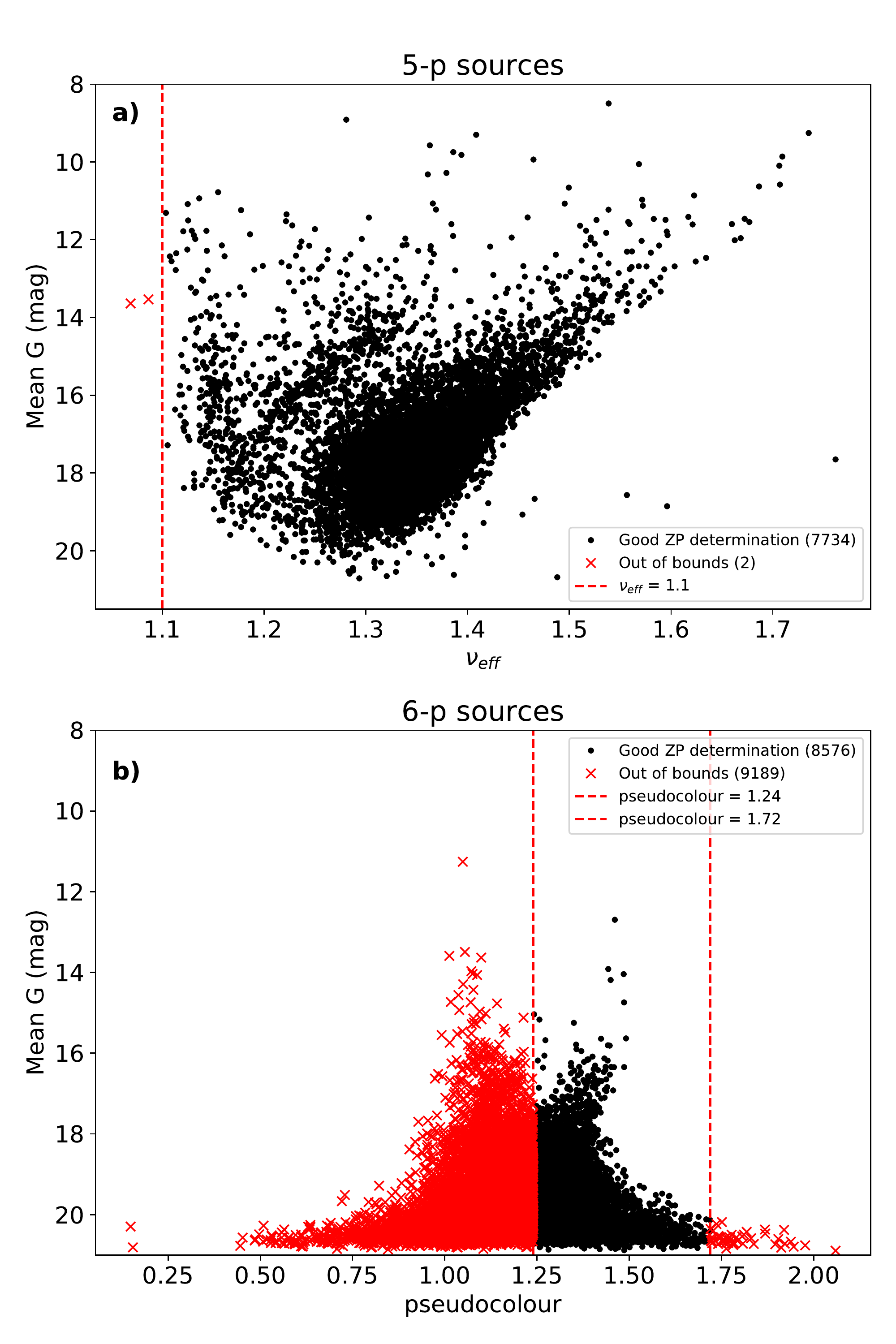}} \\[-1.0ex]
    \caption{Similar as Fig.\,\ref{fig_zeropoint_conditions} but exhibiting all the Gaia-EDR3 sources with five (5-p, panel a) and six astrometric parameters solved (6-p, panel b) within a {15\arcmin} radius centered at RA\,=\,251.767$^\circ$, Dec\,=\,$-$45.849$^\circ$. The vertical dashed lines are placed at $\nu_{\rm eff}$\,=\,1.1 (a) and pseudocolour\,=\,1.24 and 1.72 (b). } 
    \label{fig_zeropoint_condition_15arcsec}
\end{figure}

Figure\,\ref{fig_zeropoint_condition_15arcsec} presents the $G$-band magnitude versus wavenumber plot exhibiting all the 25,501 sources selected in Sect.\,\ref{sect_data}.
From the 7,736 sources with a five astrometric parameter solution (panel a), only two are located outside the $\nu_{\rm eff}$ limit (1.1-1.9).
However, $\sim$52\% of the 6-p sources (panel b) exhibit pseudocolour values outside of the (1.24-1.72) limit, indicating that about half of the 17,765 sources does not have a reliable parallax zero-point estimate.
Given that all the Wd\,1 members associated with 6-p solutions do not have reliable parallax zero-point corrections (Fig.\,\ref{fig_zeropoint_conditions}), we restricted the further analysis and results using only the 5-p sources from the Gaia-EDR3 catalogue, leading to a sample of 61 known Wd\,1 members from a total of 7,736 sources.

\subsubsection{Selection of new Wd\,1 member candidates based on the Gaia-EDR3 information}

We used the Gaia-EDR3 astrometry and photometry to select additional Wd\,1 member candidates based on the colour indices of the known members selected in Sect.\,\ref{sect_members}. The main photometric and astrometric properties of the known members of Wd\,1 are summarised in Table\,\ref{table_gaiawd1_summary}.

\begin{table}
    \centering
    \caption{General properties of the known Wd\,1 members in the Gaia-EDR3 catalogue.}
    \label{table_gaiawd1_summary}
    \begin{tabular}{l|rrr}
    \hline
Parameter  & Min &   Max &   Median\,$\pm$\,1-$\sigma$ \\
    \hline
G (mag)     &  10.94	&  17.71	&	15.18$_{-2.54}^{+1.35}$	\\
BP-RP (mag) &   3.81	&	6.87	&	4.85$_{-0.39}^{+0.61}$	\\
BP-G (mag)  &   2.32	&	5.30	&	3.26$_{-0.32}^{+0.59}$	\\
G-RP (mag)  & 	1.42	&	1.68	&	1.54$_{-0.05}^{+0.06}$	\\
RUWE        &   0.84	&	1.39	&	0.98$_{-0.06}^{+0.05}$	\\
$\pi$ (mas)$^{a}$               & 0.03	&	0.50	&	0.24$_{-0.10}^{+0.09}$	\\
$\mu_\alpha$ (mas\,yr$^{-1}$) & -2.65	&	-1.57	&	-2.27$_{-0.23}^{+0.23}$	\\
$\mu_\delta$ (mas\,yr$^{-1}$) & -4.24	&	-3.26	&	-3.69$_{-0.17}^{+0.27}$	\\
$\mu$ (mas\,yr$^{-1}$)        &  3.91	&	5.00	&	4.33$_{-0.24}^{+0.15}$	\\
    \hline
    \end{tabular} \\
    {\textbf{Notes:} The table presents the minimum, the maximum, the median and the $\pm$1-$\sigma$ deviation from the median values of each parameter.
    $(a)$ zero-point corrected parallaxes.}
\end{table}

The sample exhibits $BP-RP$ colour indices ranging between 3.8 and 6.9\,mag with a median of 4.9$_{-0.4}^{+0.6}$\,mag, suggesting a relatively high reddening towards the cluster. For completeness, the $BP-G$ colour index ranges from 2.3 to 5.3\,mag with a median of 3.3$_{-0.3}^{+0.6}$\,mag, and the $G-RP$ colour index ranges from 1.4 to 1.7\,mag with a median of 1.54$_{-0.05}^{+0.06}$\,mag. 

We defined three photometric criteria to select new candidates based on the minimum colour indices of the Wd\,1 members:
\begin{eqnarray}
BP-RP & \geq & 3.60 \\
BP-G & \geq & 2.20 \\
G-RP & \geq & 1.35
\end{eqnarray}

An additional astrometric criterion was defined by adopting a 5-$\sigma$ threshold on the two-dimensional proper motion space, filtering out objects with tangential velocities larger than $\sim$30\,{\kms} at a distance of $\sim$4.0\,kpc from the median velocity of the cluster members (see Table\,\ref{table_gaiawd1_summary}. 

The individual astrometric and photometric criteria led to samples of 1691 and 338 sources, respectively, showing that the photometric criteria are more effective for selecting the new candidates than the astrometric criterion alone.
The combination of both selection criteria led to a sample of 166 new objects.

\begin{figure}
    \centering
    {\includegraphics[width=\columnwidth]{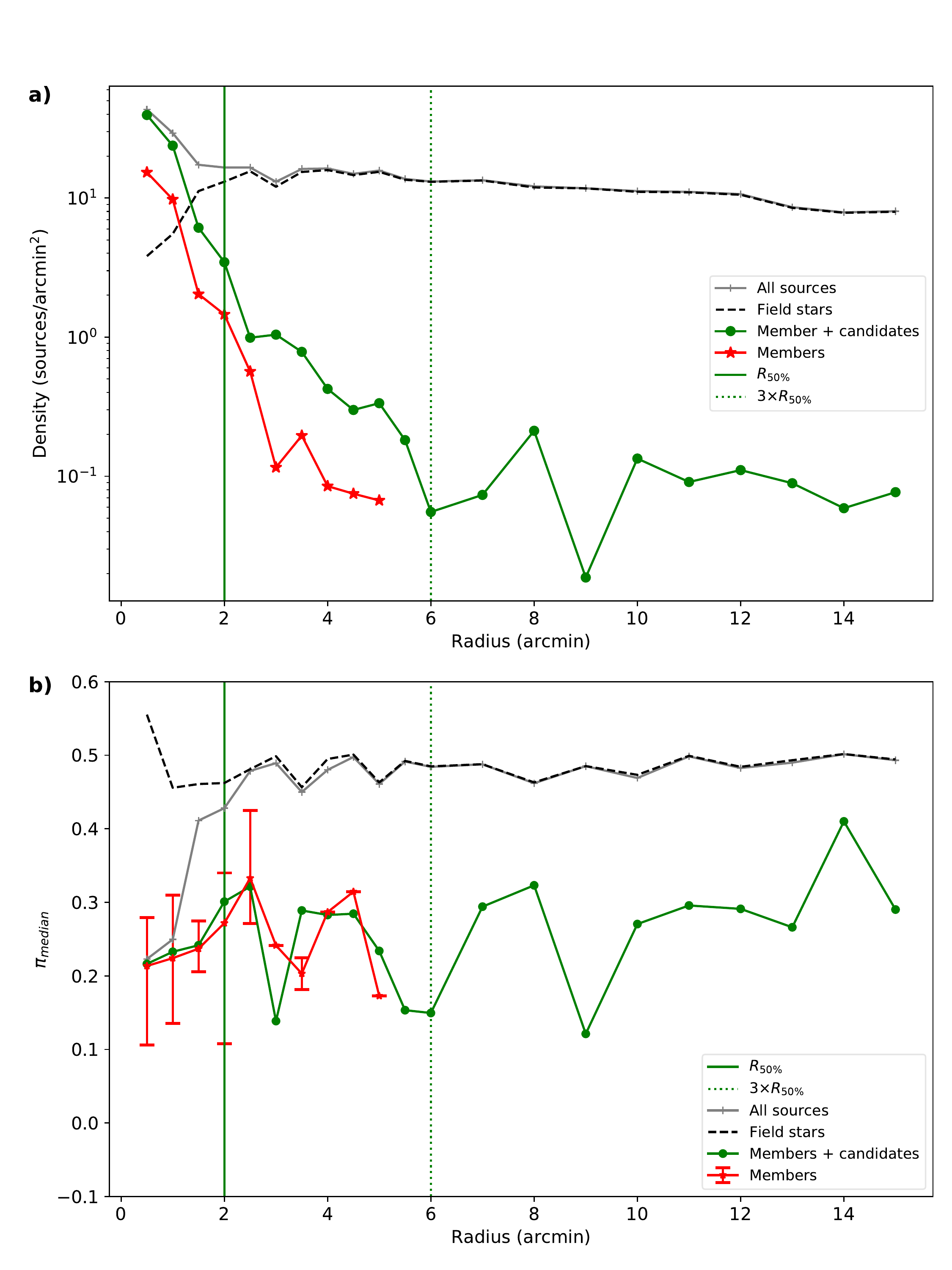}} \\[-1.0ex]
    \caption{Radial profiles for the median parallax (panel a) and stellar density (panel b) as a function of the distance to the center of the Wd\,1 cluster for the initial list of Gaia-EDR3 sources (filled grey curve and $+$ symbols), the Wd\,1 members (filled red curve and $\ast$ symbols), 
    the joint list of candidates and known members (filled green curve and circles), and the field stars (dashed black line).
    Each value was evaluated in rings of 0\farcm5 for the inner {5\arcmin} radius, and 1\farcm0 for r\,$>$\,5\arcmin. The vertical filled and dashed green lines are placed at the effective radius containing about 50\% of the cluster population (R$_{50\%}$) and at 3\,$\times$\,R$_{50\%}$.}
    \label{fig_median_parallax}
\end{figure}

We computed the radial distribution of the density of sources and the median parallax in concentric rings from the center of the Wd\,1 cluster for the distinct samples.
Figure\,\ref{fig_median_parallax}(a) shows the radial profile of the stellar density as a function of the distance from the center of the Wd\,1 cluster.
The initial list of Gaia objects (in grey) exhibits a peak at the central position of FOV (r\,=\,0\farcm5), decreasing towards larger radii.
We used the sample containing the candidates and known Wd\,1 members (in green) to estimate the effective radius of the cluster ($R_{50\%}$), defined as the radius containing about 50\% of the sample, leading to $R_{50\%}$\,=\,{2\farcm0} (shown as the vertical filled green line).
The known Wd\,1 members from \citet{Clark20} (in red) are found within the inner {5\arcmin} radius, which is about 3\,$\times$\,$R_{50\%}$\,=\,{6\farcm0} (shown as the vertical dashed green line) containing about $\sim$75\% of the cluster members (known members and candidates, shown by the green curve).

Figure\,\ref{fig_median_parallax}(b) exhibits the median parallax as a function of the distance to the center of the cluster for the same samples presented in  Fig.\,\ref{fig_median_parallax}(a). The curve containing all Gaia-EDR3 sources (filled grey curve) shows an abrupt increase of the median parallax (from $\pi$\,$\sim$\,0.20 to $\sim$0.50\,mas) in the inner {3\arcmin} radius, reaching a roughly constant value for larger radii. 
The median parallax of the full Gaia-EDR3 sample is strongly biased by the relatively large fraction of Wd\,1 members at the central region of the cluster ($r$\,$\sim$\,{0\arcmin}), while the bias diminishes towards larger distances.
The r\,$<$\,3\farcm0  limit is roughly consistent with the effective radius of the cluster defined in Fig.\,\ref{fig_median_parallax}(a), indicating that most of the stellar sources located within this radius are likely members of the cluster. 
As expected, the median parallax values for the known Wd\,1 members (red curve) are significantly smaller, assuming values ranging from 0.17 to 0.3\,mas within the inner {5\arcmin} radius. 
%
The median parallax values of the cluster members (known members and candidates, in green) exhibit a smooth increase towards distances larger than the 3\,$R_{50\%}$ limit (vertical dashed green line), suggesting the existence of another population of stellar objects with similar astrometric and photometric properties as the Wd\,1 cluster, but likely not bound to the cluster itself.
For this reason, we excluded the sources located at distances larger than {6\arcmin}, leading to a final sample of 172 sources, from which 111 are not present in the \citet{Clark20} catalogue.

For completeness, the median parallax values of the field stars (dashed black curve) exhibit a relatively constant value around 0.5\,mas within the entire range shown in the plot, indicating that the selection criteria adopted in this work were efficient to select the Wd\,1 members even at the innermost region of the cluster.

\subsection{Near-infrared photometry and far--red spectroscopy} \label{sec_phot_rsg}

Photometric $JHK_s$ observations of the Wd\,1 cluster were performed at the 60-cm Zeiss telescope of the \emph{Obsevat\'orio Pico dos Dias} (OPD/LNA, Brazil) on 2009-June-18.
To avoid saturation with the minimum exposure time, the $K_s$-band filter was replaced by a narrow filter at $\lambda$\,2.14\,{\micron} as a proxy for the $K_s$ filter. In addition to this flux reduction setup, a black cardboard mask was placed in front of the telescope, on which four holes of 2\,$\times$\,2\,cm were cut. A range of exposure times from 1 to 100\,secs were used with and without the mask to sample the bright- and faint-end stars in the field. {Additional images without mask were taken to reach a fainter magnitude limit ($K_s$\,$\sim$\,16).}

Optical spectroscopy observations ($\lambda$8400--8900{\AA}, R$\sim$12,000) of the RGB stars were taken between 2012 and 2014 at OPD, and in 2017 at the SOAR telescope (R$\sim$6,000). We used results reported by \citet{Arevalo18} for the spectral types, but we did not assign temperatures with those spectral types as the lines defining them (e.g., TiO) are not formed in the stellar photosphere \citep{davies13}.

%

\section{Results}
\label{sect_results}

\subsection{Gaia-EDR3} \label{sect_gaia}

\subsubsection{Astrometric and photometric properties of the cluster members}

\begin{figure*}
    \centering
    {\includegraphics[width=\linewidth]{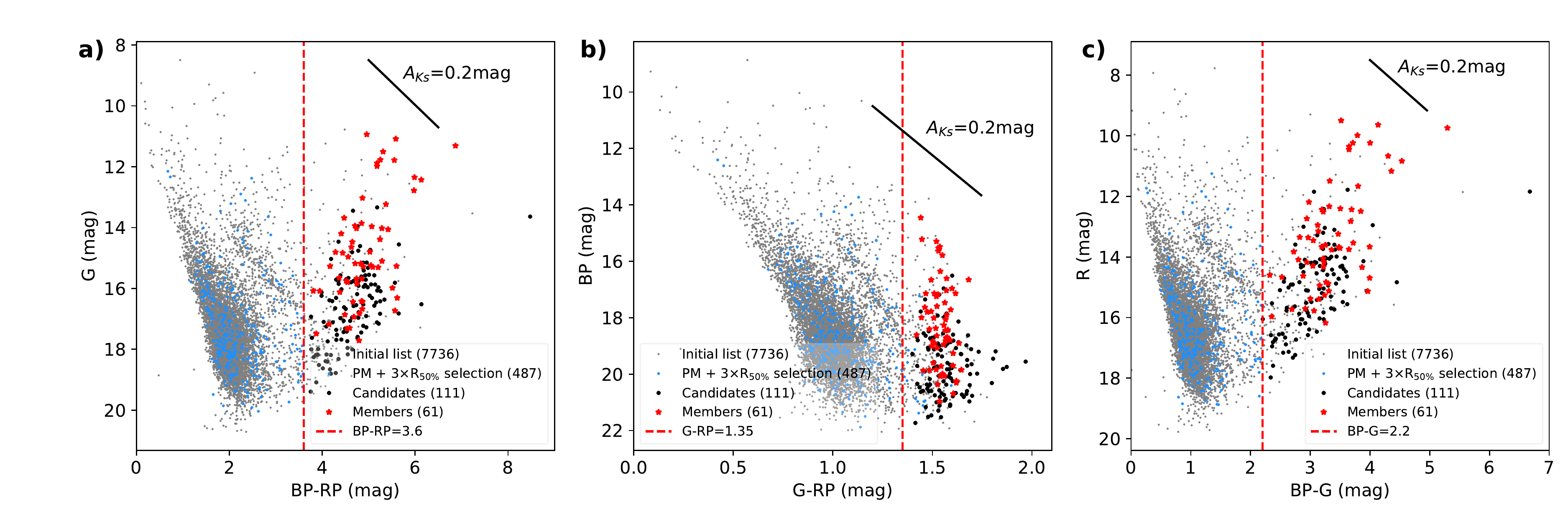}} \\[-1.0ex]
    \caption{Colour-magnitude diagrams using the photometry of Gaia-DR3 sources within a {15\arcmin} radius around the Wd\,1 central position.
    The initial list of Gaia-EDR3 sources is indicated by grey circles, known Wd\,1 members are shown as red $\ast$ symbols, sources selected using the astrometric and effective radius criteria are shown as blue circles, and the Wd\,1 member candidates are shown as black circles. The vertical dashed red lines indicate the photometric criteria adopted in this work. The black lines illustrate the reddening vectors corresponding to an $A_{\rm Ks}$\,=\,0.2\,mag using the extinction law from \citet{Damineli16}.}
    \label{fig_gaia_cmds}
\end{figure*}

Figure\,\ref{fig_gaia_cmds} presents three colour-magnitude diagrams (CMDs) showing the Gaia-EDR3 sources in the field (in grey), the known Wd\,1 members (in red), the sample selected through the astrometric and effective radius criteria only (in blue) and the sample selected using the full selection criteria (in black).
The known Wd\,1 members are located in a reddened sequence in the CMDs (shown as red $\ast$ symbols), justifying the usage of the photometric criteria (defined in Sect.\,\ref{sect_members} and indicated by the red dashed lines) to filter out a larger fraction of blue, foreground objects.
For completeness, the sample obtained with the complementary selection criteria (i.e. astrometric and radial distance selection only) still exhibits a large fraction of foreground objects.

In addition, another group of relatively bright and intermediate-reddened objects  ($G$\,$\gtrsim$\,17, 1.5\,$\leq$\,$[BP-RP]$\,$\leq$\,3.5) is located between the foreground main-sequence line and the reddened sequence corresponding to the Wd\,1 cluster. These objects are detected in all diagrams from Fig.\,\ref{fig_gaia_cmds} and they are consistent with a foreground red clump population, suggesting the existence of an evolved stellar population in the line-of-sight of the Wd\,1 cluster that could introduce a bias on the analysis if not properly taken into account.

We also plotted the reddening vectors derived from the \citet{Damineli16} extinction law in each of the CMDs from Fig\,\ref{fig_gaia_cmds}. The extinction coefficients were obtained for each Gaia filter assuming the following effective wavelengths of 532, 673, and 797\,nm for the $BP$-, $G$-, and $RP$-bands, respectively \citep{Jordi10}.

\begin{figure}
    \centering
    {\includegraphics[width=\columnwidth]{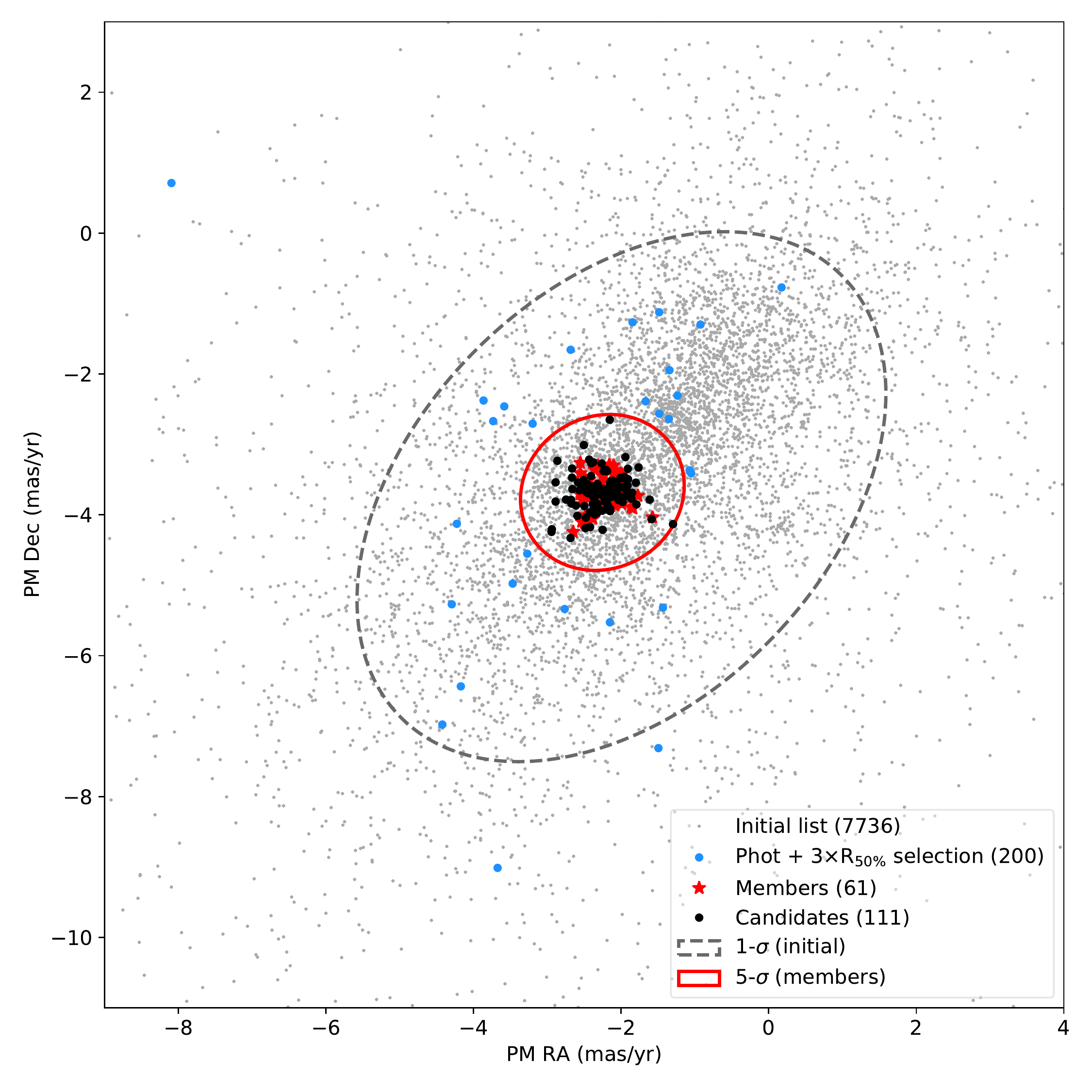}} \\[-1.0ex]
    \caption{Distribution of the proper motions of the Gaia-EDR3 sources.
    The initial list of Gaia objects are indicated by filled grey circles, the known Wd\,1 members are shown as red $\ast$ symbols, the sources selected through the photometric and effective radius criteria are shown as filled blue circles and the final sample of Wd\,1 member candidates are indicated by filled black circles.
    The 5-$\sigma$ confidence interval ellipsoid of the known Wd\,1 members (filled red curve) was used to filter out sources with large proper motions. For comparison, the 1-$\sigma$ confidence interval ellipsoid of the initial list is indicated by the dashed grey curve.}
    \label{fig_pms}
\end{figure}

%
%
Figure\,\ref{fig_pms} exhibits the distribution of each sample in the proper motion space.
The proper motion of the known Wd\,1 members (in red) range $-$2.65\,$\leq$\,$\mu_\alpha$\,$\leq$\,$-$1.57\,mas\,yr$^{-1}$ (median of $-$2.26$_{-0.23}^{+0.28}$\,mas\,yr$^{-1}$) and $-$4.24\,$\leq$\,$\mu_\delta$\,$\leq$\,$-$2.39\,mas\,yr$^{-1}$ (median of $-$3.69$_{-0.17}^{+0.24}$\,mas\,yr$^{-1}$).
The 1-$\sigma$ values correspond to a velocity dispersion of about $\pm$\,5.\,\kms at a distance of $\sim$\,4\,kpc.
The 5-$\sigma$ confidence interval ellipsoid for the proper motions of the known Wd\,1 members, shown as the red curve in Fig.\,\ref{fig_pms}, was used as the astrometric selection criterion defined in Sect.\,\ref{sect_members}. 
The ellipsoid is based on the covariance matrix and the linear Pearson correlation coefficient of the two-dimensional data (that is, $\mu_\alpha$ versus $\mu_\delta$), and the size of the ellipsoid is defined as the number of standard deviations ($\sigma$) of the distribution. 

The known Wd\,1 members (in red) exhibit a compact distribution in terms of proper motions, as expected for a cluster of stars. The Wd\,1 members also lie inside the 1-$\sigma$ confidence interval of the initial list (indicated by the dashed grey curve). The center of the distribution of the members is offset by ($-$0.49, $-$0.29)\,mas from the center of the initial list of Gaia-EDR3 sources, corresponding to a linear projected velocity of about $\sim$\,10\,\kms at a distance of $\sim$\,4.0\,kpc.
About half of the sample selected through the photometric and angular distance criteria defined in Sect.\,\ref{sect_members} are located within the 5-$\sigma$ confidence interval of the known members (dashed red curve), leading to the final list of 111 new Wd\,1 member candidates.

\begin{figure}
    \centering
    {\includegraphics[width=\linewidth]{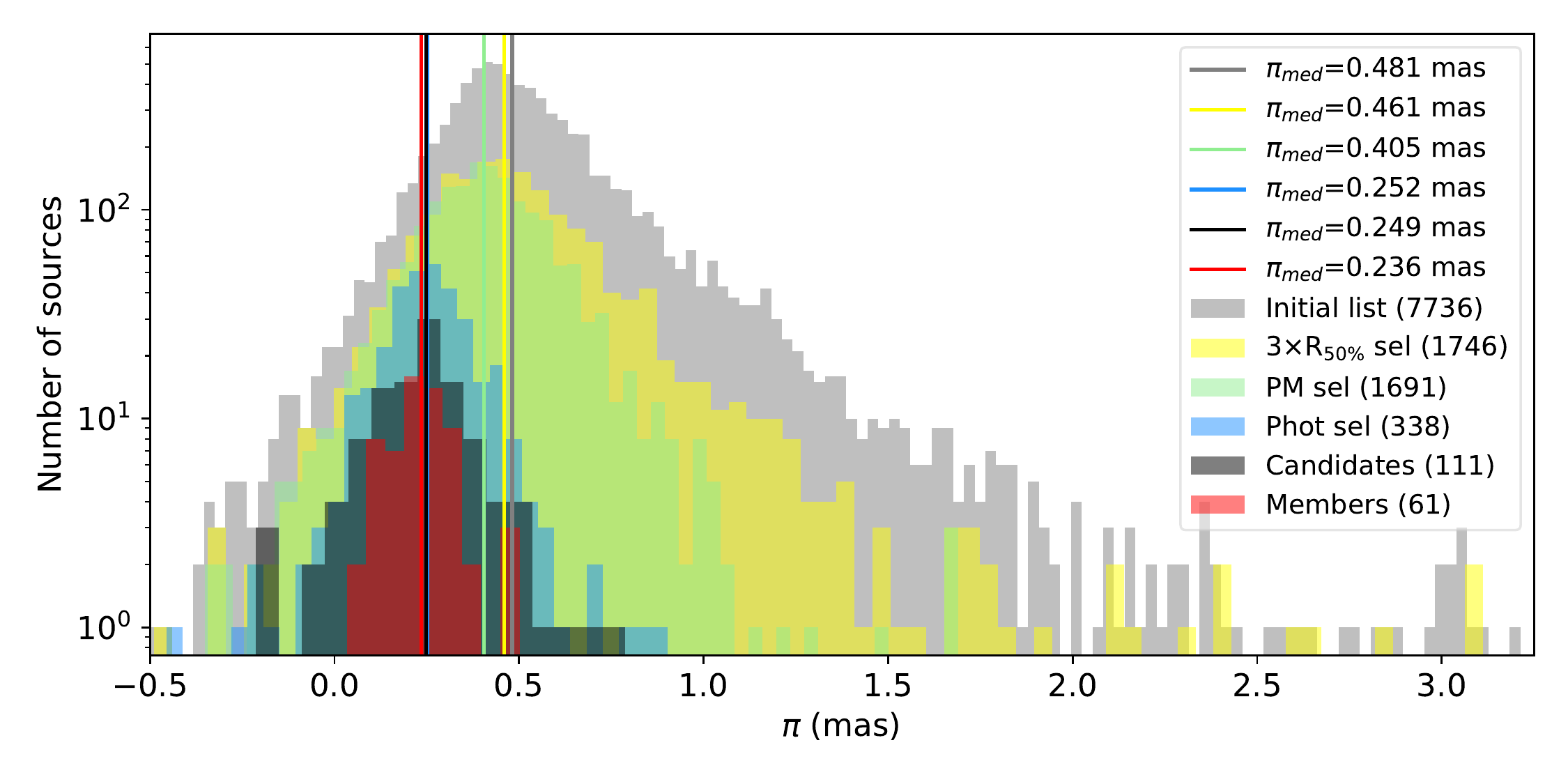}} \\[-2.0ex]
    \caption{Distribution of the parallax values for the Gaia-DR3 sources.
    The initial Gaia-EDR3 sample is indicated as grey, known Wd\,1 members are shown in red, sources selected through effective radius, astrometric, and photometric criteria are shown as yellow, green and blue, respectively. The final sample of member candidates is indicated in black. The vertical lines are placed at the median parallax values of each sample.}
    \label{fig_parallax_histogram}
\end{figure}

Figure\,\ref{fig_parallax_histogram} presents the distribution of the parallax values for the initial list of Gaia-EDR3 sources (in grey), the known Wd\,1 members (in red), the samples selected through angular distance and effective radius (yellow), astrometric (green) and photometric criteria (blue), and the known cluster members + candidates (in black).
As expected, the selection criteria applied to the Gaia-EDR3 sources was efficient to filter out objects associated with large parallax values (e.g. at closer distances), leading to a final sample exhibiting a relatively narrow distribution of parallaxes with a median value around 0.25\,mas (indicated by the vertical black line). The resulting distribution is in good agreement with the value observed for the known Wd\,1 members (0.236\,mas, shown as the vertical red line).

\begin{figure*}
    \centering
    {\includegraphics[width=\linewidth]{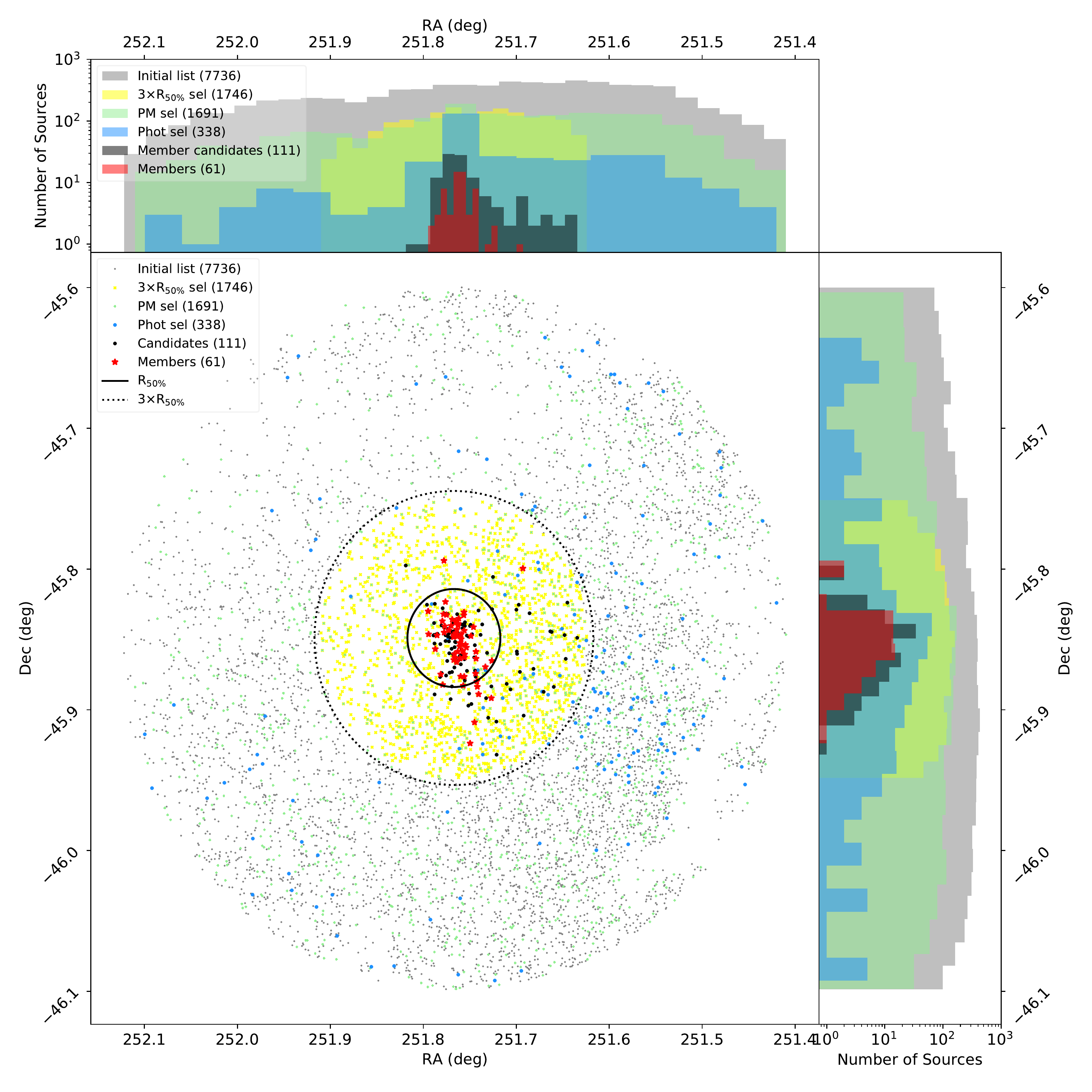}} \\[-1.0ex]
    \caption{Spatial distribution of the Gaia-EDR3 sources within a radius of {15\arcmin} centered at the Wd\,1 position (RA\,=\,251.76707$^\circ$, Dec\,=\,$-$45.84899$^\circ$).
    The initial list of Gaia objects are indicated by grey circles,
    the known Wd\,1 members are shown as red $\ast$ symbols, and the Wd\,1 member candidates are indicated by filled black circles.
    The sub-samples selected through angular distance, astrometric and photometric criteria individually are shown as yellow $\times$, green $+$, and blue $\times$ symbols, respectively. The top and right panels show the distribution of the same sources as a function of the right ascension and declination, respectively.
    The effective radius containing 50\% of the cluster population ($R_{50\%}$\,=\,2\arcmin) and the 3\,$\times$\,$R_{50\%}$ radius are indicated by the filled and dotted black curves, respectively.}
    \label{fig_fov}
\end{figure*}

Figure\,\ref{fig_fov} exhibits the spatial distribution of the sources in the FOV. As expected, the Wd\,1 members and the new candidate Wd\,1 members (shown as red $ast$ and black circle symbols, respectively) exhibit a cusp distribution around the central position of Wd\,1 in both axis.
The sources selected through the photometric criteria alone (in blue) exhibits a tailed distribution extending to {$\sim$12\arcmin} to the West direction of Wd\,1, while about 20 sources are located at r\,$>$\,{10\arcmin} the SE direction of the cluster. These objects exhibit similar colours as the cluster members and are likely located at the same distance (i.e. the same spiral arm) as the cluster itself.

\subsubsection{Distance inference from the Gaia-EDR3 parallaxes}
\label{sec_distance_gaia_mean}

We followed the methodology of \citet{CantatGaudin18} to infer the distance to the Wd\,1 cluster based on the parallax values ($\pi$) and their uncertainties ($\sigma_\pi$) of selected members, computing the distance to the cluster ($d$) through a maximum likelihood procedure defined as:
\begin{equation}
    \prod_{i=1}^{n} P(\pi_i | d, \sigma_{\pi_i}) = \prod_{i=1}^{n} \frac{1}{\sqrt{2 \pi \sigma_{\pi_i}^2}} \exp{\left( -\frac{(\pi_i - 1/d)^2}{2\sigma_{\pi_i}^2} \right)}
    \label{eq_likelihood}
\end{equation}

\begin{figure}
    \centering
    {\includegraphics[width=\columnwidth]{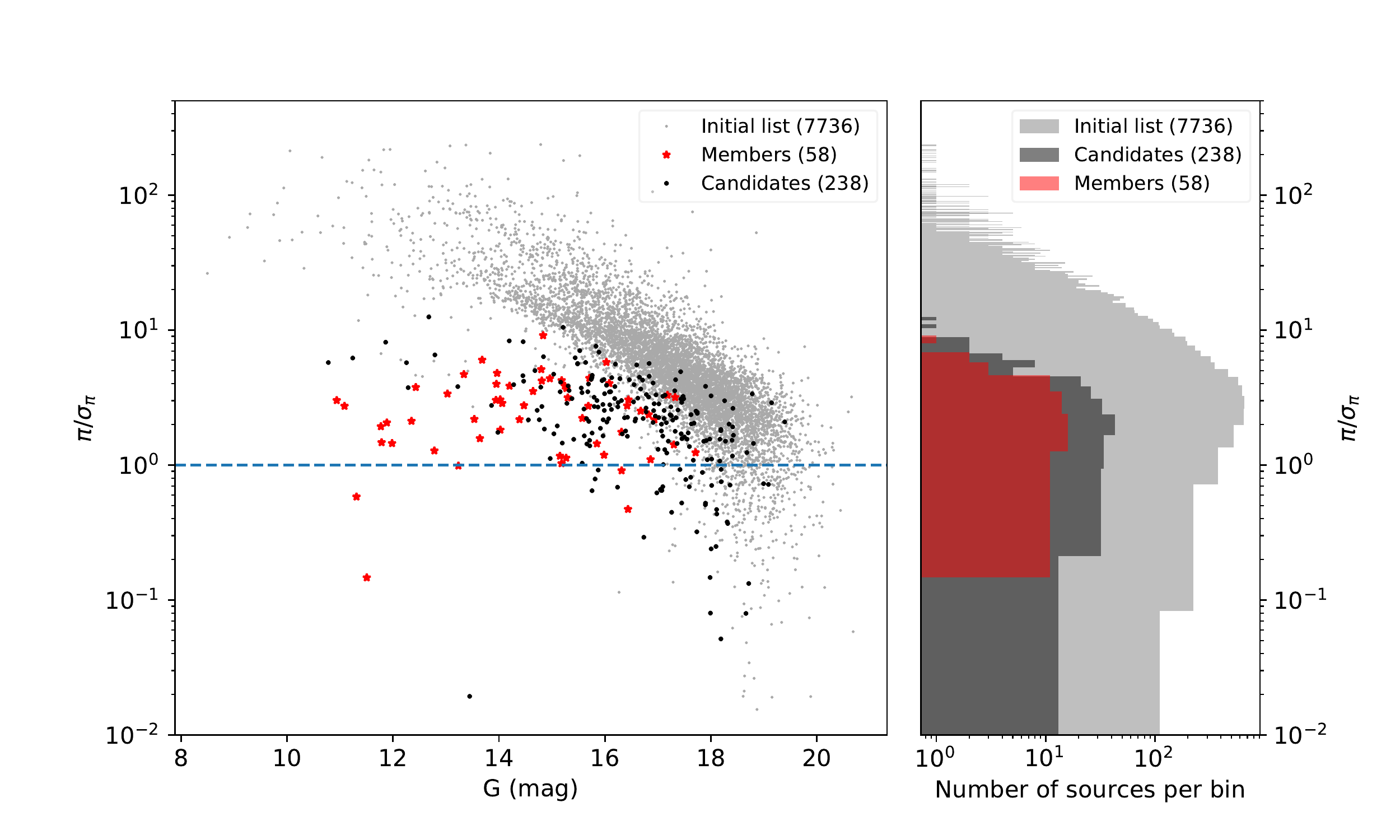}} \\[-1.0ex]
    \caption{Left panel: Distribution of the parallax-to-error ratio as a function of the $G$-band magnitude of the Gaia sources.
    The initial list of Gaia objects are indicated by grey circles, the known Wd\,1 members are shown as red $\ast$ symbols, the new member candidates are indicated by the black circles. The horizontal dashed blue line indicates y\,=\,1.
    Right panel: distribution of the parallax-to-error ratio values for the three samples. The number of sources per bin increases from left to right.}
    \label{fig_parallaxerrors}
\end{figure}

As pointed out by those authors, this method does not take into account the physical depth of the cluster neither the position of the sources within the cluster, assuming that all members are located at the same distance.
This approximation is true for distant clusters, from which the physical depth is relatively small when compared to the individual parallax uncertainties of its members.
To verify if such approximation is valid for Wd\,1, Fig.\,\ref{fig_parallaxerrors} shows the ratio between the parallax and their errors ($\pi/\sigma_\pi$) as a function of the $G$-band magnitude of the Gaia-EDR3 sources.
The plot exhibits a large fraction of the selected Wd\,1 sources (red $\ast$ symbols) associated with relatively small parallax-to-error ratio values (0.01\,$\lesssim$\,$\pi/\sigma_\pi$\,$\lesssim$\,10) when compared to the values observed for the whole sample of Gaia EDR3 sources in the field (grey points, with $\pi/\sigma_\pi$ values up to $\sim$\,10$^2$).
These findings confirm the assumption of large uncertainties on the individual parallaxes of the cluster members and, therefore, its distance can be inferred using Eq.\,\ref{eq_likelihood}.

\begin{figure}
    \centering
    {\includegraphics[width=\columnwidth]{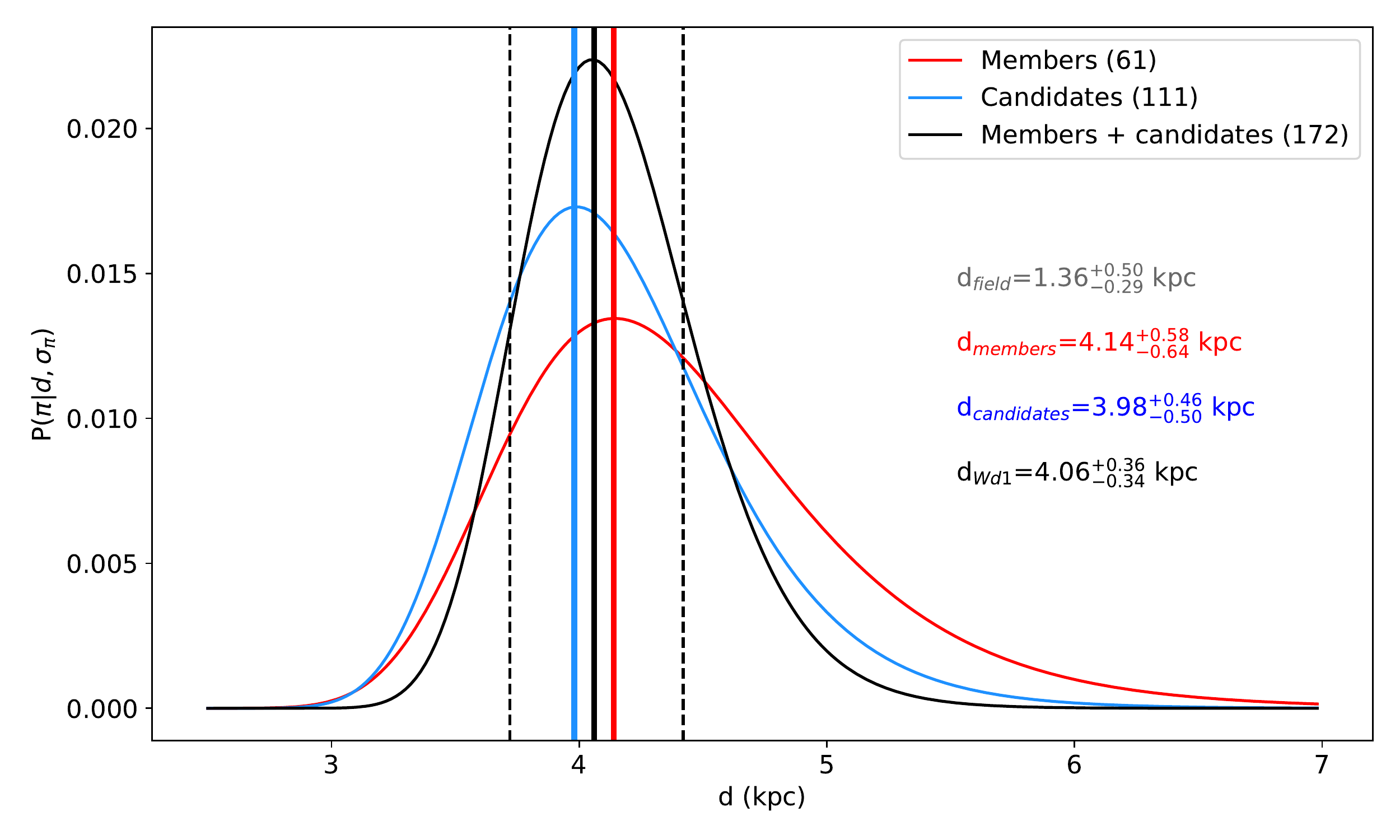}} \\[-1.0ex]
    \caption{Probability density function of the posterior distance to Wd\,1 based on the known members of the cluster (in red), the selected candidates (in blue), and the {union of both samples} (in black). The most likely distance of the final sample is 4.06$^{+0.36}_{-0.34}$\,kpc, indicated by the filled black line, and the 1-$\sigma$ errors (68\%) are shown as dashed black lines. The distances based on the other samples are indicated by the vertical filled lines. The distances are listed in their corresponding colours and the distance to the field-stars (not shown in the plot) is listed in grey.}
    \label{fig_all_pdf}
\end{figure}

The result of Eq.\,\eqref{eq_likelihood} is a probability density function (PDF), peaking at the distance where the product $P(\pi | d, \sigma_{\pi})$ assumes its maximum value. 
The resulting PDF considering the 58 known Wd\,1 members is shown as the red curve in Fig.\,\ref{fig_all_pdf}. 
The PDF exhibits an asymmetric profile, with an elongated tail towards large distances. The PDF peaks at the distance of  4.14\,kpc$^{+0.60}_{-0.66}$\,kpc, with the upper and lower uncertainties defined as the 68\% confidence interval, consistent with a 1-$\sigma$ estimate from other methods.

The determination of the distance to the cluster and its associated errors was improved by considering the new Wd\,1 member candidates, as shown in Fig.\,\ref{fig_all_pdf}. Overlaid on the PDF of the known Wd\,1 members (in red), the plot also shows the PDF considering only the new candidates (in blue), and the union of both samples (members + candidates, in black). The sample of 172 objects led to a more precise inference of the distance to the cluster, estimated as $d_{\rm gedr3}$\,$=$\,4.06$^{+0.36}_{-0.34}$\,kpc.  
 
For completeness, we computed the weighted-mean parallax of the field stars to infer their mean distance. The weighted mean parallax ($\langle \pi \rangle$) was evaluated using Eqs.\,(1)-(3) from \citet{Navarete19}, leading to $\langle \pi \rangle$\,=\,0.74\,$\pm$\,0.20\,mas. We estimated the distance of the field stars by simply inverting their mean parallax and propagating the errors, leading to a mean distance of 1.36$^{+0.50}_{-0.29}$\,kpc. Such distance is consistent with the location of the near-side of the Carinae arm at the line-of-sight of Wd\,1 cluster \citet{Reid19}.

A study of the W36 eclipsing binary system (Paper\,II) led to a independent distance estimate of
$d_{\rm w36}$\,=\,4.03\,$\pm$\,0.25\,kpc, in agreement with our Gaia-EDR3 results.
From here on we adopt the weighted mean distance to the cluster as $d_{\rm wd1}$\,=\,4.05\,$\pm$\,0.20\,kpc (further discussions are presented in Sect.\,\ref{sect_discussion_distance}).

\subsection{Luminosity and Age of the Red Supergiants} \label{subsect_age}

The improved accuracy of extinction and the distance obtained in this work has a direct impact on the determination of the luminosities of the cluster members. As a consequence, the parameters of the cluster as a whole can also be inferred with greater confidence.

{The photometric results for the four RSGs and six YHGs are listed in Table\,\ref{table_RSGages}. 
The photon statistics accuracy is $\sim$0.01\,mag for the bright members (degrading for the fainter members), but due to the relatively small number of 2MASS calibrating sources in the FOV, the photometric accuracy was limited to $\sim$\,0.05\,mag in the bright-end ($K_s$\,$<$\,13). Using the references in the caption of  Table\,\ref{table_RSGages} we derived the intrinsic colours and temperatures of our targets. We adopted the  \citet{Damineli16} reddening law  to get the $AK_s$ extinction: $AK_s$\,$=$\,0.449$E_{J-K_s}$ and $AK_s$\,$=$\,1.300$E_{H-K_s}$ (column 6 in their Table\,\ref{table_RSGages}).  
The average extinction of the RSGs and YHGs $A_{Ks}$\,$=$\,0.80\,$\pm$\,0.14 is compatible with that derived from the OB stars ($A_{Ks}$\,$=$\,0.74$\pm$0.08, D16), indicating that these evolved cool stars do not have significant circumstellar extinction. The exception is W75 (in addition to W8a) which has a significantly larger than the average extinction, as suspected by \citet{beasor21}.}

{The total luminosity of each RSG (column\,7 of Table\,\ref{table_RSGages}) was calculated by using the bolometric calibration suggested by \citet{Davies2018} -- BC$_K$\,$=$\,3.00\,$\pm$\,0.18 -- and the mean distance of $d_{\rm wd1}$\,$=$\,4.05\,kpc from Sect.\,\ref{sec_distance_gaia_mean}.
The range of RSGs luminosities is relatively narrow ($\Delta$\,(log(L/$\rm{L}_\odot$)\,$=$\,0.24) indicating a narrow range of RSG progenitor masses.}
Column 8 of Table\,\ref{table_RSGages} shows luminosities from \citet{beasor21} (their Table\,4, column\, 6), scaled with  our distance of $d$\,$=$\,4.05\,kpc. The results are in excellent agreement  in luminosity indicating that the differences between the two works mostly arise from the $AK_s$ extinction towards individual targets.

\setlength{\tabcolsep}{9pt}
\begin{table*}
    \centering
    \caption{Classification, $JHK_s$ photometry, reddening and bolometric luminosities for Red Supergiants (first four) and Yellow Hypergiants (last six).}
    \label{table_RSGages}
    \begin{tabular}{llccccccc}
    \hline
ID  & Spectral Type & $\log(T_{\rm eff})$ &$J$ & $H$ & $K_s$ & $A_{Ks}$ &  $\log(L/\rm{L}_\odot)^a$& B21$^b$ \\ 
 \hline
\multicolumn{2}{l}{Red Supergiants:} \\
W237 & M3-4\,I  & - & 4.86 & 3.33 & 2.44 & 0.67\,$\pm$\,0.16 & 5.21\,$\pm\,$0.03 & 5.30 \\
W20  & M3-4\,I  & -  & 6.28 & 4.33 & 3.24 & 0.95\,$\pm$\,0.17 & 4.98\,$\pm\,$0.03 & 5.16 \\
W75  & M1\,Ia  & -  & 6.96 & 4.68 & 3.44 & 1.23\,$\pm$\,0.15 & 5.00\,$\pm\,$0.03 & 4.90 \\
W26  & M1-1.5\,Ia& - & 4.92 & 3.16 & 2.40 & 0.68\,$\pm$\,0.02 & 5.20\,$\pm$\,0.03 & 5.45 \\
\hline
\multicolumn{2}{l}{Yellow Hypergiants:} \\
W4   & F3\,Ia+ &3.81  & 6.06 & 4.92 & 4.45 & 0.58\,$\pm$\,0.15 & 4.96\,$\pm\,$0.03 & 5.04 \\
W8a  & F8\,Ia+ & 3.72 & 6.50 & 5.41 & 3.82 & 1.05\,$\pm$\,0.20 & 5.17\,$\pm\,$0.03 & 4.73 \\
W32  & F5\,Ia+ &3.79 & 5.69 & 4.49 & 3.87 & 0.72\,$\pm$\,0.17 & 5.17\,$\pm\,$0.03 & 5.21 \\
W16a & A5\,Ia+ &3.91  & 6.55 & 5.37 & 4.78 & 0.74\,$\pm$\,0.16 & 5.28\,$\pm\,$0.03 & 4.95 \\
W12a & F1\,Ia+ & 3.83  & 6.73 & 5.50 & 4.84 & 0.79\,$\pm$\,0.18 & 4.97\,$\pm\,$0.03 & 4.88 \\
W265 & F1-5\,Ia+ &3.82 & 6.74 & 5.45 & 4.75 & 0.82\,$\pm$\,0.16 & 4.94\,$\pm\,$0.03 & 4.88 \\
    \hline
    \end{tabular} \\
    {\textbf{Notes:} $(a)$ Spectral Types of RSGs and YHGs from \citet{Arevalo18} and \citet{Clark20}, respectively, $BC_K$\,$=$\,3.00  \citep{Davies2018}, NIR colours \citep{koornneef83};
     $BC_V$ \citep{flower96}. 
     $(b)$ Bolometric luminosities from \citet[][Table\,4]{beasor21} scaled to \textbf{4.05\,kpc}}.
\end{table*}

\setlength{\tabcolsep}{6pt}

Table\,\ref{table_RSGages} also displays our photometric results for the six YHGs. The bolometric correction follows \citet{flower96}. Our results are in very good agreement with those reported by \citet{beasor21}, which is based on independent photometry.

We adopted the \emph{Binary Population and Spectral Synthesis} method \citep[BPASS,][]{eldridge20} for estimating the age based on the mean of the luminosity of the four RSGs (interpolated into their Z\,$=$\,0.020 models, column\,6 of Table\,1). The mean RSG luminosity, $\log(L_{\rm RSG}/\rm{L}_\odot)$\,$=$\,5.10\,$\pm$\,0.11 corresponds to a cluster age of 10.7\,$\pm$\,1\,Myr.

\section{Discussion}
\label{sect_discussion}

\subsection{The distance to Wd1 confirmed by independent methods}
\label{sect_discussion_distance}

We used the Gaia-EDR3 analysis to infer the distance to the Westerlund\,1 cluster based on the sample of evolved, high-mass stars from \citet{Clark20}. We further included additional member candidates selected through astrometric and photometric criteria to obtain a final sample of 172 sources with Gaia-EDR3 counterparts, leading to the distance of $d_{\rm gedr3}$\,=\,4.06$^{+0.36}_{-0.34}$\,kpc. 

We also found that the sources in \citet{Clark20} are relatively red objects, as expected for a cluster located at large distances in the Galactic Plane. The main impact of that fact is regarding the parallax zero-point derivation. For the range of brightness and colours of the cluster members in the Gaia-EDR3 catalogue, we found that the parallax zero-point correction is well-determined for Gaia objects with five astrometric parameters solution (5p), but not for objects with six astrometric parameters solution (6p, see Figs.\,\ref{fig_zeropoint_conditions}  and \ref{fig_zeropoint_condition_15arcsec}). Therefore, we excluded the 6p sources from our analysis. This criterion was fundamental for removing any systematic biases due to dubious parallax zero-points, a critical correction for objects at relatively large distances.

Our Gaia-EDR3 distance to Wd\,1 is in excellent agreement with the distance reported by \citet{beasor21}, 4.12$^{+0.66}_{-0.33}$\,kpc.
{The major difference between their results and ours is mainly related to the sample selection and the adoption of fine-tuning procedures in this work for excluding sources with unreliable zero-point parallax correction (see Sect.\,\ref{sect_zp}).} 

\citet{Negueruela22} reported an independent confirmation of the distance to Wd\,1 also based on the OB stellar population from \citet{Clark20}.
They reported eight distance estimates (see their Table\,1), adopting case C as the distance to the cluster, 4.23$^{+0.23}_{-0.21}$\,kpc.
Interestingly, cases F and H of their Table\,1 are very similar to our results (their source selection criteria for case H likely match the same as adopted by us).
However, apparently they chose case C as an intermediate value of the eight cases instead, stating that their distance estimate is robust and independent in terms of their source selection methodology, the parallax zero-point correction or choice of sources with 5- or 6-parameters astrometric solutions.
Taking into account the independent distance of W36 in Paper\,II ($d_{\rm w36}$\,=\,4.03\,$\pm$\,0.25\,kpc), the closest distances reported in their Table\,1 (cases F, H) are likely more reliable.
They adopted a distinct parallax zero-point correction and a specific distance prior for the distribution of OB stars in the Galactic plane (see references therein), which led to a distance in agreement with our results and with those from \citet{beasor21}.

{These three Gaia-EDR3 distances indicate that the Wd\,1 cluster is located between 4.0 and 4.3\,kpc with unprecedented accuracy. In addition, the Gaia-EDR3 distances can be confirmed by independent measurements,} such as the modelling of eclipsing binary systems.
Indeed, Paper\,II presents the modelling of the eclipsing binary system W36, obtaining a distance of {$d_{\rm w36}$\,$=$\,4.03$\pm$0.25\,kpc}. The distance to W36 is consistent within 1-$\sigma$ with {each of the three independent Gaia-EDR3 distances, being an important result once} the parallax zero-point corrections are uncertain in the Gaia-EDR3 for {most of} objects associated with very red colour indices \citep{Lindegren21b}, {as likely observed for the bulk of} the Wd\,1 cluster members {(see Sect.\,\ref{sect_zp})}.

{Table\,\ref{table_wd1distances} summarises the distances reported by \citet{beasor21}, \citet{Negueruela22}, Paper\,I (this work), and also includes the distance to the massive binary system W36 from Paper\,II.}

\setlength{\tabcolsep}{3pt}
\begin{table}
    \centering
    \caption{{Gaia-EDR3 distances to Wd\,1 reported in the literature and results obtained in Paper\,I (this work) and Paper\,II.}}
    \label{table_wd1distances}
    \begin{tabular}{ccl}
        \hline
        \hline
        Method    & Distance (kpc)         & Reference \\
        \hline
        Gaia-EDR3 & 4.12$^{+0.66}_{-0.36}$ & \citet{beasor21}     \\
        Gaia-EDR3 & 4.23$^{+0.23}_{-0.21}$ & \citet{Negueruela22} \\
        Gaia-EDR3 & 4.06$^{+0.36}_{-0.34}$ & Paper\,I (this work) \\
        W36 ecl. binary & 4.03$\pm$0.25 & Paper\,II            \\
        \hline
    \end{tabular}
\end{table}
\setlength{\tabcolsep}{6pt}

At the distance of 4.05$\pm$0.20\,kpc, the effective size of Wd\,1, $r$\,=\,2\farcm0 (Figs.\,\ref{fig_median_parallax} and \ref{fig_fov}) corresponds to a linear radius of 2.33\,$\pm$\,0.12\,pc, and our sample is located within three times the effective radius, 7.0\,$\pm$\,0.4\,pc.
It is worth mentioning that our sample is biased towards the high-mass stellar content of the cluster. In general, the low-mass counterparts are likely less gravitationally bound to the cluster, extending to larger distances from the center of the cluster.
Both the diameter of the cluster and the low-density halo concept are in agreement with the analysis presented by \citet{Negueruela22}. 

Wd\,1 is located at the galactic latitude of $\ell$\,=\,339\fdg55.
Based on recent Galactic structure results \citep{Reid19}, the cluster is located in the line-of-sight of three major galactic spiral arms: the Carina arm at a distance of 0.5\,kpc, the Scutum-Crux near arm at 3.5\,kpc, and the Norma near arm at 5.5\,kpc. Our results suggest that the distance of Westerlund\,1 is consistent with the {position of the far side} of the Scutum-Crux near arm, {or even the near side of the Norma arm}. {\citet{Negueruela22} provide strong arguments favouring that the cluster is located within the Norma arm, but further studies are still necessary to confirm the true association of the cluster with one or another spiral arm.}

\subsection{The Age of Westerlund 1}

We further compare our results on the age of Wd\,1 with those from \citet{beasor21}. Those authors tested the robustness of different methods for evaluating the age of the cluster by using a wide range of stellar populations (RSGs, PMS and the eclipsing binary W13). They reported the age of 7.2$^{+1.1}_{-2.3}$\,Myr for the pre-MS population, while the faintest RSG led to a older age between 9.2-11.7\,Myr. Our results for the age of the RSGs based on BPASS models (11\,$\pm$\,1\,Myr, see Sect.\,\ref{subsect_age}) is in excellent agreement with those authors.

In paper\,II, we derived the age of the eclipsing binary W36 (using \citet{yusof22} evolutionary models) as $\sim$6\,Myr. Only the W36B component has a reliable age estimation, since the peeling of external layers of the W36A component shifts its effective temperature to the blue, mimicking an earlier age. Furthermore, 
the impact of mass accretion onto W36B is low since the mass lost by W36A is mainly through its stellar wind, so the W36B stellar parameters are little affected. The age of W36B is not far from the age limit of W13 ($<$\,5\,Myr) reported by \citet{beasor21} and also to $\sim$\,7\,Myr suggested by \citet{Hosek18}. Compared to the ages of the RSGs, the much younger W36B and W13 sources are evidence of -- at least -- two episodes of star formation in the cluster.

\section{Conclusions}
\label{sect_summary}

Our astrometric and photometric analysis using the Gaia Early Data Release 3 (Gaia-EDR3) resulted in the distance of $d_{\rm gedr3}$\,$=$\,4.06$^{+0.36}_{-0.34}$\,kpc to the massive stellar cluster Westerlund\,1.
The agreement between the Gaia-EDR3 distance and the distance of the W36 eclipsing binary system $d_{\rm w36}$\,$=$\,4.03$\pm$\,0.25\,kpc (Paper\,II) validates our results and, in particular, the parallax zero-point determination for the reddest Wd\,1 members classified as 5-p sources in the Gaia-EDR3 catalogue. The weighted mean distance of both independent estimates led to $d_{\rm wd1}$\,=\,4.05\,$\pm$\,0.20\,kpc, with a unprecedented error of 5\%.

We reported the age of the RSG stars using BPASS models for the average luminosity as 9.7--11.7\,Myr, in good agreement with that of W75 reported by \citet[][9.2--11.7\,Myr]{beasor21} based on a broader range of approaches and a different dataset.

The age of W36B is $\sim$\,6.5\,Myr,  is close to the  $<$\,5\,Myr reported by \citet{beasor21} for the binary system W13. These two works indicate a range of star formation in Wd1 of 5--12\,Myr, in line with the synthetic cluster calculated by \citet{yusof22} combining rotating and non-rotating models with an age spread of 5--10\,Myr. \citet{yusof22} showed that their result was able to reproduce qualitatively the observed population of RSGs, YHGs, and WRs in Wd\,1. 

\section*{Acknowledgements}

The work of FN is supported by NOIRLab, which is managed by the Association of Universities for Research in Astronomy (AURA) under a cooperative agreement with the National Science Foundation.
FN and AD thanks to Funda\c{c}\~{a}o de Amparo \`a Pesquisa do Estado de S\~{a}o Paulo (FAPESP) for support through process number 2017/19181-9 (FN) and 2019/02029-2+2011/51680-6 (AD) and to CNPq 301490/2019-8 (AD).

We thank B. Davies for careful reading of our manuscript, and for productive questions/remarks which has improved our manuscript.
Based in part on observations carried out at Observat\'orio do Pico dos Dias (OPD), which is operated by LNA/MCTI, Brazil.
Based  in part on observations obtained at the Southern Astrophysical Research (SOAR) telescope, which is a joint project of the Minist\'{e}rio da Ci\^{e}ncia, Tecnologia e Inova\c{c}\~{o}es (MCTI/LNA) do Brasil, the US National Science Foundation’s NOIRLab, the University of North Carolina at Chapel Hill (UNC), and Michigan State University (MSU).

\section*{Data availability}

The data sets were derived from observations available in the public domain: \url{https://www.cosmos.esa.int/web/gaia/earlydr3}.

Tables\,\ref{table_RSGages}, \ref{table_Wd1members_Gaia5p}, \ref{table_Wd1members_Gaia6p} and \ref{table_Wd1members_noGaia} are also available in the CDS.

\bibliographystyle{mnras}


\appendix

\section{Full tables}

\begin{onecolumn}
\begin{landscape}
\setlength{\tabcolsep}{5pt}
\begin{table*}
\caption{List of known Westerlund 1 members with Gaia EDR3 counterparts associated with a 5-parameter astrometric solution (5-p). This table is available online at the CDS.}
\label{table_Wd1members_Gaia5p}
\begin{tabular}{llcccccccccccc}
\hline
\multicolumn{1}{c}{Name} & 
\multicolumn{1}{c}{GSC} &  \multicolumn{1}{c}{2MASS} &  \multicolumn{1}{c}{Gaia EDR3} &  \multicolumn{1}{c}{RUWE} &  \multicolumn{1}{c}{$\pi$} &   \multicolumn{1}{c}{$\pi_{\rm ZP}$} &  \multicolumn{1}{c}{$\mu_\alpha$} &  \multicolumn{1}{c}{$\mu_\delta$} &  \multicolumn{1}{c}{$\nu_{\rm eff}$} &  \multicolumn{1}{c}{min($\sigma_{\rm phot}$)} &  \multicolumn{1}{c}{G} &  \multicolumn{1}{c}{BP} &  \multicolumn{1}{c}{RP} \\
\multicolumn{1}{c}{} & 
\multicolumn{1}{c}{designation} &  \multicolumn{1}{c}{designation} &  \multicolumn{1}{c}{designation} &  \multicolumn{1}{c}{} &  \multicolumn{1}{c}{(mas)} &   \multicolumn{1}{c}{(mas)} &  \multicolumn{1}{c}{(mas\,yr$^{-1}$)} &  \multicolumn{1}{c}{(mas\,yr$^{-1}$)} &  \multicolumn{1}{c}{(\micron$^{-1}$)} &  \multicolumn{1}{c}{($\sigma$)} &  \multicolumn{1}{c}{(mag)} &  \multicolumn{1}{c}{(mag)} &  \multicolumn{1}{c}{(mag)} \\
\hline
W2a	&	--	&	16465971-4550513	&	5940106758703247360	&	0.94	&	0.264\,$\pm$\,0.044	&	$-$0.068	&	$-$2.015\,$\pm$\,0.053	&	$-$3.414\,$\pm$\,0.045	&	1.156	&	115.2	&	13.7	&	16.7	&	12.2	\\
W4	&	S8UV052685	&	16470142-4550373	&	5940106763014985088	&	0.96	&	0.180\,$\pm$\,0.059	&	$-$0.045	&	$-$2.237\,$\pm$\,0.073	&	$-$3.415\,$\pm$\,0.063	&	1.136	&	99.5	&	10.9	&	14.5	&	9.5	\\
W5	&	2MIU38SH	&	16470298-4550199	&	5940106797374726784	&	1.09	&	0.154\,$\pm$\,0.056	&	$-$0.073	&	$-$2.495\,$\pm$\,0.068	&	$-$3.467\,$\pm$\,0.059	&	1.151	&	87.8	&	14.5	&	17.6	&	12.9	\\
W6a	&	2MIU38SN	&	16470303-4550235	&	5940106797374726144	&	1.01	&	0.249\,$\pm$\,0.059	&	$-$0.076	&	$-$2.222\,$\pm$\,0.071	&	$-$3.803\,$\pm$\,0.062	&	1.150	&	68.9	&	15.2	&	18.4	&	13.6	\\
W7	&	S8UV052690	&	16470363-4550144	&	5940106793062992128	&	0.89	&	0.122\,$\pm$\,0.059	&	$-$0.014	&	$-$2.379\,$\pm$\,0.074	&	$-$3.713\,$\pm$\,0.063	&	1.131	&	109.2	&	11.9	&	15.5	&	10.4	\\
W8a	&	S8UV052686	&	16470480-4550251	&	5940106041460479488	&	0.88	&	0.119\,$\pm$\,0.061	&	$-$0.018	&	$-$2.376\,$\pm$\,0.075	&	$-$3.786\,$\pm$\,0.065	&	1.129	&	134.1	&	11.8	&	15.5	&	10.2	\\
W10	&	S8UV052706	&	16470334-4550346	&	5940106797366404480	&	1.00	&	0.137\,$\pm$\,0.063	&	$-$0.075	&	$-$1.989\,$\pm$\,0.075	&	$-$3.421\,$\pm$\,0.065	&	1.133	&	80.6	&	14.4	&	18.1	&	12.8	\\
W11	&	S8UV052697	&	--	&	5940106758703254656	&	0.95	&	0.144\,$\pm$\,0.052	&	$-$0.071	&	$-$2.480\,$\pm$\,0.064	&	$-$4.007\,$\pm$\,0.056	&	1.143	&	144.8	&	13.9	&	17.2	&	12.3	\\
W12a	&	S8UV052692	&	16470222-4550590	&	5940106763006638848	&	1.00	&	0.154\,$\pm$\,0.073	&	$-$0.006	&	$-$2.493\,$\pm$\,0.088	&	$-$3.692\,$\pm$\,0.076	&	1.114	&	68.1	&	12.3	&	16.7	&	10.7	\\
W13	&	S8UV052698	&	16470646-4550261	&	5940106037157928064	&	0.94	&	0.094\,$\pm$\,0.052	&	$-$0.071	&	$-$2.497\,$\pm$\,0.069	&	$-$4.112\,$\pm$\,0.060	&	1.147	&	79.4	&	14.0	&	17.2	&	12.5	\\
W16a	&	S8UV052691	&	16470661-4550422	&	5940106041460473216	&	0.89	&	0.097\,$\pm$\,0.066	&	$-$0.018	&	$-$2.130\,$\pm$\,0.086	&	$-$3.760\,$\pm$\,0.072	&	1.121	&	71.1	&	11.8	&	15.8	&	10.2	\\
W21	&	2MIU38V7	&	16470110-4551135	&	5940106659935758976	&	0.96	&	0.296\,$\pm$\,0.058	&	$-$0.077	&	$-$2.517\,$\pm$\,0.081	&	$-$3.582\,$\pm$\,0.062	&	1.137	&	55.1	&	14.8	&	18.3	&	13.2	\\
W23a	&	2MIU38UU	&	16470256-4551088	&	5940106758703255168	&	0.99	&	0.163\,$\pm$\,0.054	&	$-$0.074	&	$-$2.317\,$\pm$\,0.072	&	$-$3.298\,$\pm$\,0.058	&	1.132	&	78.0	&	14.0	&	17.7	&	12.4	\\
W24	&	2MIU38V4	&	16470215-4551126	&	5940106758709795712	&	0.90	&	0.212\,$\pm$\,0.055	&	$-$0.079	&	$-$2.367\,$\pm$\,0.073	&	$-$3.630\,$\pm$\,0.060	&	1.141	&	61.5	&	15.2	&	18.7	&	13.7	\\
W26	&	S8UV052688	&	16470540-4550367	&	5940106041452150272	&	1.09	&	0.053\,$\pm$\,0.092	&	$-$0.024	&	$-$1.841\,$\pm$\,0.127	&	$-$3.909\,$\pm$\,0.108	&	1.104	&	21.7	&	11.3	&	16.6	&	9.7	\\
W32	&	--	&	16470369-4550435	&	5940106763006654208	&	1.00	&	0.190\,$\pm$\,0.070	&	$-$0.034	&	$-$2.262\,$\pm$\,0.087	&	$-$3.615\,$\pm$\,0.074	&	1.125	&	119.8	&	11.1	&	15.2	&	9.6	\\
W33	&	S8UV052689	&	16470413-4550485	&	5940106002789009536	&	0.84	&	0.085\,$\pm$\,0.059	&	$-$0.011	&	$-$2.053\,$\pm$\,0.078	&	$-$3.728\,$\pm$\,0.063	&	1.132	&	131.5	&	12.0	&	15.6	&	10.5	\\
W35	&	2MIU38TW	&	16470417-4550533	&	5940106007100789760	&	1.03	&	0.201\,$\pm$\,0.064	&	$-$0.078	&	$-$2.088\,$\pm$\,0.086	&	$-$3.862\,$\pm$\,0.069	&	1.143	&	56.6	&	15.3	&	18.8	&	13.7	\\
W42a	&	S8UV052694	&	16470325-4550522	&	5940106763006644096	&	0.90	&	0.092\,$\pm$\,0.072	&	$-$0.018	&	$-$1.970\,$\pm$\,0.091	&	$-$3.502\,$\pm$\,0.078	&	1.113	&	85.2	&	12.8	&	17.1	&	11.2	\\
W47	&	2MIU38VF	&	16470260-4551177	&	5940106763014970624	&	1.32	&	0.203\,$\pm$\,0.074	&	$-$0.078	&	$-$2.649\,$\pm$\,0.094	&	$-$4.242\,$\pm$\,0.079	&	1.154	&	39.2	&	15.7	&	18.9	&	14.1	\\
W52	&	2MIU397M	&	16470184-4551294	&	5940106659927373056	&	1.03	&	0.269\,$\pm$\,0.056	&	$-$0.072	&	$-$2.140\,$\pm$\,0.070	&	$-$3.873\,$\pm$\,0.060	&	1.138	&	134.0	&	14.0	&	17.5	&	12.4	\\
W54	&	S8UV051192	&	--	&	5940106007092338176	&	1.05	&	0.131\,$\pm$\,0.067	&	$-$0.080	&	$-$2.244\,$\pm$\,0.083	&	$-$3.801\,$\pm$\,0.069	&	1.133	&	56.1	&	15.1	&	18.8	&	13.5	\\
W55	&	S8UV051040	&	16465841-4551313	&	5940106655624021632	&	0.99	&	0.172\,$\pm$\,0.049	&	$-$0.073	&	$-$2.333\,$\pm$\,0.064	&	$-$3.849\,$\pm$\,0.052	&	1.152	&	103.9	&	14.6	&	17.8	&	13.1	\\
W57a	&	S8UV051189	&	16470136-4551456	&	5940106655624023552	&	0.95	&	0.203\,$\pm$\,0.060	&	$-$0.051	&	$-$2.251\,$\pm$\,0.075	&	$-$3.448\,$\pm$\,0.061	&	1.139	&	43.7	&	13.0	&	16.4	&	11.5	\\
W60	&	S8UV050868	&	--	&	5940105904013098496	&	0.97	&	0.157\,$\pm$\,0.055	&	$-$0.078	&	$-$2.141\,$\pm$\,0.070	&	$-$3.658\,$\pm$\,0.058	&	1.143	&	48.7	&	15.3	&	18.6	&	13.8	\\
W63a	&	S8UV050867	&	--	&	5940105904021499776	&	0.97	&	0.118\,$\pm$\,0.054	&	$-$0.076	&	$-$2.061\,$\pm$\,0.067	&	$-$3.376\,$\pm$\,0.056	&	1.163	&	37.5	&	15.7	&	18.5	&	14.1	\\
W65	&	S8UV050871	&	--	&	5940105904013103744	&	0.86	&	0.196\,$\pm$\,0.056	&	$-$0.079	&	$-$2.445\,$\pm$\,0.071	&	$-$3.760\,$\pm$\,0.058	&	1.148	&	33.1	&	15.7	&	18.9	&	14.2	\\
W66	&	2MIU3984	&	16470396-4551377	&	5940106007100719872	&	1.03	&	0.101\,$\pm$\,0.085	&	$-$0.078	&	$-$2.552\,$\pm$\,0.108	&	$-$3.262\,$\pm$\,0.090	&	1.119	&	17.5	&	16.0	&	19.8	&	14.3	\\
W71	&	S8UV052000	&	16470846-4550493	&	5940106037148742272	&	0.88	&	0.065\,$\pm$\,0.066	&	$-$0.069	&	$-$2.384\,$\pm$\,0.084	&	$-$4.054\,$\pm$\,0.081	&	1.128	&	97.2	&	13.2	&	17.0	&	11.7	\\
W74	&	S8UV052684	&	16470708-4550130	&	5940199877892578816	&	0.96	&	0.061\,$\pm$\,0.060	&	$-$0.077	&	$-$2.534\,$\pm$\,0.075	&	$-$4.097\,$\pm$\,0.070	&	1.149	&	68.0	&	15.2	&	18.4	&	13.7	\\
W78	&	S8UV053443	&	16470155-4549580	&	5940106797374731008	&	0.96	&	0.148\,$\pm$\,0.049	&	$-$0.071	&	$-$1.984\,$\pm$\,0.060	&	$-$3.474\,$\pm$\,0.052	&	1.149	&	123.9	&	13.9	&	17.1	&	12.4	\\
W84	&	S8UV052342	&	16465904-4550284	&	5940106763014989184	&	1.00	&	0.177\,$\pm$\,0.048	&	$-$0.073	&	$-$2.544\,$\pm$\,0.056	&	$-$3.655\,$\pm$\,0.052	&	1.170	&	97.1	&	15.3	&	18.0	&	13.8	\\
W228b	&	2MIU39FQ	&	16465803-4553008	&	5940106247618875648	&	1.01	&	0.264\,$\pm$\,0.065	&	$-$0.060	&	$-$2.266\,$\pm$\,0.084	&	$-$3.645\,$\pm$\,0.063	&	1.173	&	7.6	&	16.1	&	18.4	&	14.6	\\
W232	&	S8UV049673	&	16470144-4552351	&	5940105869661755904	&	1.02	&	0.224\,$\pm$\,0.053	&	$-$0.072	&	$-$2.348\,$\pm$\,0.067	&	$-$3.694\,$\pm$\,0.053	&	1.165	&	108.3	&	14.8	&	17.6	&	13.3	\\
W243	&	S8UV049907	&	16470749-4552290	&	5940105830990286208	&	1.15	&	0.012\,$\pm$\,0.081	&	$-$0.022	&	$-$1.574\,$\pm$\,0.108	&	$-$4.036\,$\pm$\,0.086	&	1.125	&	102.6	&	11.5	&	15.3	&	10.0	\\
W265	&	S8UV053674	&	16470627-4549238	&	5940199877886921728	&	0.98	&	0.293\,$\pm$\,0.078	&	$-$0.008	&	$-$2.431\,$\pm$\,0.096	&	$-$3.500\,$\pm$\,0.087	&	1.107	&	80.7	&	12.4	&	17.0	&	10.8	\\
W373	&	2MIU2A8Q	&	16465773-4553200	&	5940106247618872960	&	1.00	&	0.171\,$\pm$\,0.044	&	$-$0.070	&	$-$2.502\,$\pm$\,0.056	&	$-$3.797\,$\pm$\,0.047	&	1.160	&	126.1	&	14.2	&	17.1	&	12.7	\\
WR I	&	2MIU38VN	&	16470088-4551206	&	5940106655630578560	&	0.94	&	0.151\,$\pm$\,0.086	&	$-$0.064	&	$-$1.956\,$\pm$\,0.122	&	$-$3.679\,$\pm$\,0.097	&	1.122	&	15.4	&	16.3	&	20.3	&	14.7	\\
WR N	&	S8UU115458	&	16465989-4555255	&	5940105354265570688	&	1.02	&	0.090\,$\pm$\,0.079	&	$-$0.083	&	$-$1.773\,$\pm$\,0.103	&	$-$3.730\,$\pm$\,0.081	&	1.120	&	33.7	&	15.3	&	19.3	&	13.7	\\
WR T	&	S8UV055302	&	16464628-4547582	&	5940107106612387968	&	0.95	&	0.240\,$\pm$\,0.055	&	$-$0.075	&	$-$2.528\,$\pm$\,0.066	&	$-$3.772\,$\pm$\,0.059	&	1.153	&	107.9	&	15.0	&	18.0	&	13.5	\\
1005	&	S8UV050530	&	--	&	5940106689983753728	&	1.05	&	0.265\,$\pm$\,0.069	&	$-$0.080	&	$-$2.355\,$\pm$\,0.089	&	$-$3.679\,$\pm$\,0.075	&	1.135	&	50.3	&	15.3	&	19.0	&	13.7	\\
\hline
\end{tabular} \\
{\textbf{Notes:} The columns are as follows:
(1) Name of the source (the coordinates are listed in \citet{Clark19}); 
(2) designation of the source in the GSC 2.3 catalogue;
(3) designation of the source in the 2MASS PSC catalogue \citep{Skrutskie06};
(4) Identification of the Gaia EDR3 source;
(5) Renormalized Unit Weight Error;
(6) Parallax listed from the Gaia EDR3 and its error;
(7) Parallax zero-point correction derived from \citet{Lindegren21b};
(8)-(9) Proper motion in RA and Decl axis and their error;
(10) $\nu_{\rm eff}$;
(11) Minimum photometric error of the G, BP and RP-bands (in $\sigma$ units);
(12)-(14) Magnitude in the G, BP and RP-bands, respectively.}
\end{table*}

\setcounter{table}{0}
\begin{table}
\caption{}
\contcaption{}
\label{table_Wd1members_Gaia5p-cont}
\begin{tabular}{llcccccccccccc}
\hline		
\multicolumn{1}{c}{Name} & 
\multicolumn{1}{c}{GSC} &  \multicolumn{1}{c}{2MASS} &  \multicolumn{1}{c}{Gaia EDR3} &  \multicolumn{1}{c}{RUWE} &  \multicolumn{1}{c}{$\pi$} &   \multicolumn{1}{c}{$\pi_{\rm ZP}$} &  \multicolumn{1}{c}{$\mu_\alpha$} &  \multicolumn{1}{c}{$\mu_\delta$} &  \multicolumn{1}{c}{$\nu_{\rm eff}$} &  \multicolumn{1}{c}{min($\sigma_{\rm phot}$)} &  \multicolumn{1}{c}{G} &  \multicolumn{1}{c}{BP} &  \multicolumn{1}{c}{RP} \\
\multicolumn{1}{c}{} &
\multicolumn{1}{c}{designation} &  \multicolumn{1}{c}{designation} &  \multicolumn{1}{c}{designation} &  \multicolumn{1}{c}{} &  \multicolumn{1}{c}{(mas)} &   \multicolumn{1}{c}{(mas)} &  \multicolumn{1}{c}{(mas\,yr$^{-1}$)} &  \multicolumn{1}{c}{(mas\,yr$^{-1}$)} &  \multicolumn{1}{c}{(\micron$^{-1}$)} &  \multicolumn{1}{c}{($\sigma$)} &  \multicolumn{1}{c}{(mag)} &  \multicolumn{1}{c}{(mag)} &  \multicolumn{1}{c}{(mag)} \\
\hline
1006	&	--	&	16465443-4553300	&	5940106208947392640	&	0.98	&	0.113\,$\pm$\,0.102	&	$-$0.059	&	$-$2.353\,$\pm$\,0.129	&	$-$3.734\,$\pm$\,0.103	&	1.164	&	14.3	&	16.9	&	19.9	&	15.4	\\
1010	&	S8UV050162	&	--	&	5940106625576008320	&	1.39	&	0.443\,$\pm$\,0.143	&	$-$0.055	&	$-$1.991\,$\pm$\,0.188	&	$-$3.844\,$\pm$\,0.153	&	1.179	&	15.7	&	17.5	&	19.8	&	16.0	\\
1016	&	S8UV049398	&	16465817-4552470	&	5940106243307139840	&	0.97	&	0.285\,$\pm$\,0.086	&	$-$0.056	&	$-$2.103\,$\pm$\,0.107	&	$-$3.639\,$\pm$\,0.087	&	1.175	&	27.7	&	17.2	&	19.9	&	15.7	\\
1019	&	2MIU398W	&	16465837-4551488	&	5940106621270836096	&	0.98	&	0.144\,$\pm$\,0.101	&	$-$0.060	&	$-$2.305\,$\pm$\,0.121	&	$-$3.770\,$\pm$\,0.101	&	1.153	&	18.8	&	17.3	&	20.4	&	15.8	\\
1020	&	S8UV049864	&	--	&	5940106625576002944	&	0.97	&	0.189\,$\pm$\,0.054	&	$-$0.059	&	$-$2.552\,$\pm$\,0.067	&	$-$3.410\,$\pm$\,0.056	&	1.181	&	70.9	&	16.1	&	18.6	&	14.7	\\
1021	&	S8UU115692	&	16465877-4554319	&	5940105453031539456	&	1.01	&	0.226\,$\pm$\,0.090	&	$-$0.061	&	$-$2.160\,$\pm$\,0.115	&	$-$3.290\,$\pm$\,0.090	&	1.147	&	21.6	&	16.7	&	20.0	&	15.1	\\
1028	&	2MIU398A	&	16470132-4551385	&	5940106659927364480	&	0.97	&	0.318\,$\pm$\,0.100	&	$-$0.059	&	$-$2.102\,$\pm$\,0.124	&	$-$3.298\,$\pm$\,0.100	&	1.158	&	9.5	&	17.3	&	20.2	&	15.7	\\
1029	&	2MIU38JE	&	16470150-4549502	&	5940106793069533568	&	1.02	&	0.217\,$\pm$\,0.092	&	$-$0.062	&	$-$2.362\,$\pm$\,0.114	&	$-$3.358\,$\pm$\,0.096	&	1.144	&	12.7	&	16.8	&	20.1	&	15.3	\\
1030	&	S8UV049321	&	16470167-4552580	&	5940105869661752192	&	0.95	&	0.429\,$\pm$\,0.047	&	$-$0.073	&	$-$2.465\,$\pm$\,0.059	&	$-$3.726\,$\pm$\,0.049	&	1.159	&	138.5	&	14.8	&	17.8	&	13.4	\\
1033	&	S8UV049698	&	16470235-4552340	&	5940105869661755648	&	1.02	&	0.100\,$\pm$\,0.069	&	$-$0.080	&	$-$2.262\,$\pm$\,0.085	&	$-$3.878\,$\pm$\,0.069	&	1.148	&	61.7	&	15.8	&	19.1	&	14.3	\\
1034	&	S8UV051193	&	16470254-4551488	&	5940105904021502976	&	0.98	&	0.194\,$\pm$\,0.061	&	$-$0.078	&	$-$2.236\,$\pm$\,0.076	&	$-$3.615\,$\pm$\,0.063	&	1.156	&	41.9	&	15.8	&	18.8	&	14.3	\\
1037	&	S8UV052778	&	16470285-4550066	&	5940106797374729344	&	0.88	&	0.188\,$\pm$\,0.068	&	$-$0.062	&	$-$2.272\,$\pm$\,0.085	&	$-$3.539\,$\pm$\,0.071	&	1.144	&	14.7	&	16.4	&	19.7	&	14.9	\\
1038	&	S8UV054160	&	--	&	5940200629515963136	&	1.04	&	0.340\,$\pm$\,0.181	&	$-$0.026	&	$-$2.933\,$\pm$\,0.245	&	$-$3.355\,$\pm$\,0.183	&	1.303	&	38.5	&	18.3	&	19.5	&	17.2	\\
1040	&	S8UV052764	&	--	&	5940106797374728576	&	1.22	&	0.391\,$\pm$\,0.089	&	$-$0.079	&	$-$2.103\,$\pm$\,0.110	&	$-$3.409\,$\pm$\,0.091	&	1.149	&	38.6	&	15.7	&	18.7	&	14.2	\\
1046	&	S8UV053170	&	--	&	5940199873591749632	&	0.92	&	0.168\,$\pm$\,0.065	&	$-$0.060	&	$-$2.063\,$\pm$\,0.080	&	$-$3.611\,$\pm$\,0.073	&	1.157	&	27.6	&	16.1	&	19.0	&	14.6	\\
1049	&	S8UV055733	&	16470667-4547384	&	5940200771245787008	&	0.95	&	0.161\,$\pm$\,0.056	&	$-$0.074	&	$-$2.344\,$\pm$\,0.069	&	$-$3.785\,$\pm$\,0.060	&	1.128	&	76.1	&	14.1	&	17.9	&	12.5	\\
1050	&	S8UV053039	&	16470677-4549554	&	5940199877886909696	&	0.92	&	0.226\,$\pm$\,0.074	&	$-$0.061	&	$-$2.369\,$\pm$\,0.090	&	$-$3.688\,$\pm$\,0.081	&	1.150	&	23.8	&	16.4	&	19.6	&	14.9	\\
1052	&	S8UV049235	&	16470699-4552560	&	5940105835302011776	&	1.03	&	0.168\,$\pm$\,0.135	&	$-$0.055	&	$-$2.552\,$\pm$\,0.175	&	$-$3.711\,$\pm$\,0.135	&	1.147	&	11.3	&	17.7	&	21.0	&	16.2	\\
1058	&	S8UV146458	&	--	&	5940105972740981120	&	0.96	&	0.029\,$\pm$\,0.098	&	$-$0.065	&	$-$1.863\,$\pm$\,0.129	&	$-$3.893\,$\pm$\,0.110	&	1.123	&	10.9	&	16.7	&	20.7	&	15.1	\\
1062	&	--	&	16471063-4550465	&	5940199809167415680	&	0.99	&	0.189\,$\pm$\,0.087	&	$-$0.061	&	$-$2.103\,$\pm$\,0.107	&	$-$3.831\,$\pm$\,0.088	&	1.151	&	16.4	&	16.9	&	20.1	&	15.4	\\
1063	&	--	&	16471074-4549476	&	5940199907951488640	&	1.00	&	0.037\,$\pm$\,0.079	&	$-$0.061	&	$-$2.220\,$\pm$\,0.093	&	$-$3.777\,$\pm$\,0.080	&	1.148	&	21.8	&	16.4	&	19.7	&	14.8	\\
W19	&	--	&	16470485-4550593	&	5940106007100730368	&	1.51	&	$-$0.052\,$\pm$\,0.086	&	--	&	$-$2.50\,$\pm$\,0.11	&	$-$3.90\,$\pm$\,0.09	&	1.138	&	67.1	&	14.4	&	18.1	&	12.7	\\
W56a	&	S8UV050674	&	16465894-4551486	&	5940106621264278784	&	1.59	&	0.130\,$\pm$\,0.082	&	--	&	$-$2.38\,$\pm$\,0.10	&	$-$3.44\,$\pm$\,0.09	&	1.144	&	94.9	&	14.1	&	17.4	&	12.5	\\
W238	&	S8UV049975	&	--	&	5940105904021494144	&	3.62	&	0.496\,$\pm$\,0.194	&	--	&	$-$3.26\,$\pm$\,0.24	&	$-$4.90\,$\pm$\,0.19	&	1.150	&	76.9	&	14.5	&	17.5	&	12.9	\\
W28	&	S8UV052693	&	--	&	5940106037148748416	&	4.50	&	$-$0.232\,$\pm$\,0.168	&	--	&	$-$2.05\,$\pm$\,0.22	&	$-$2.99\,$\pm$\,0.20	&	1.149	&	123.9	&	13.6	&	16.8	&	12.0	\\
W241	&	S8UV050263	&	--	&	5940105899709768192	&	4.79	&	0.914\,$\pm$\,0.25	&	--	&	$-$6.87\,$\pm$\,0.32	&	$-$5.08\,$\pm$\,0.24	&	1.150	&	68.3	&	15.1	&	18.2	&	13.4	\\
\hline
\end{tabular}
\end{table}

\setlength{\tabcolsep}{6pt}
\end{landscape}
\end{onecolumn}
\begin{onecolumn}
\begin{landscape}
\setlength{\tabcolsep}{5pt}
\begin{table*}
\caption{List of known Westerlund 1 members with Gaia-EDR3 counterparts associated with a 6-parameter astrometric solution (6-p). This table is available online at the CDS.}
\label{table_Wd1members_Gaia6p}
\begin{tabular}{llccccccccccc}
\hline
\multicolumn{1}{c}{Name} &   \multicolumn{1}{c}{GSC} &  \multicolumn{1}{c}{2MASS} &  \multicolumn{1}{c}{Gaia EDR3} &  \multicolumn{1}{c}{RUWE} &  \multicolumn{1}{c}{$\pi$} &  \multicolumn{1}{c}{$\mu_\alpha$} &  \multicolumn{1}{c}{$\mu_\delta$} &  \multicolumn{1}{c}{Pseudocolour} &  \multicolumn{1}{c}{min($\sigma_{\rm phot}$)} &  \multicolumn{1}{c}{G} &  \multicolumn{1}{c}{BP} &  \multicolumn{1}{c}{RP} \\
\multicolumn{1}{c}{} & \multicolumn{1}{c}{designation} &  \multicolumn{1}{c}{designation} &  \multicolumn{1}{c}{designation} &  \multicolumn{1}{c}{} &  \multicolumn{1}{c}{(mas)} &   \multicolumn{1}{c}{(mas\,yr$^{-1}$)} &  \multicolumn{1}{c}{(mas\,yr$^{-1}$)} &  \multicolumn{1}{c}{(\micron$^{-1}$)} &  \multicolumn{1}{c}{($\sigma$)} &  \multicolumn{1}{c}{(mag)} &  \multicolumn{1}{c}{(mag)} &  \multicolumn{1}{c}{(mag)} \\
\hline
W20	&	S8UV050219	&	16470309-4552189	&	5940105904023386752	&	1.19	&	0.337\,$\pm$\,0.124	&	$-$3.085\,$\pm$\,0.148	&	$-$4.119\,$\pm$\,0.119	&	1.049	&	30.0	&	11.3	&	16.9	&	9.7	\\		
W25	&	S8UV112186	&	--	&	5940106041452309376	&	0.74	&	0.061\,$\pm$\,0.169	&	$-$1.911\,$\pm$\,0.255	&	$-$4.050\,$\pm$\,0.195	&	1.043	&	11.2	&	16.7	&	19.3	&	13.9	\\		
W49	&	S8UV112174	&	--	&	5940106763006673664	&	1.02	&	$-$0.089\,$\pm$\,0.475	&	$-$3.200\,$\pm$\,0.615	&	$-$2.892\,$\pm$\,0.436	&	1.317	&	111.3	&	19.4	&	--	&	--	\\		
W62a	&	2MIU3988	&	16470248-4551380	&	5940106659935750144	&	0.98	&	$-$0.253\,$\pm$\,0.165	&	$-$2.019\,$\pm$\,0.631	&	$-$3.361\,$\pm$\,0.402	&	1.115	&	260.1	&	16.8	&	--	&	--	\\		
WR W	&	2MIU3H42	&	16470761-4549222	&	5940199907954832768	&	1.07	&	0.062\,$\pm$\,0.177	&	$-$2.077\,$\pm$\,0.217	&	$-$3.614\,$\pm$\,0.179	&	1.068	&	7.0	&	18.1	&	21.5	&	16.5	\\		
WR X	&	--	&	16471413-4548320	&	5940199976670849152	&	1.17	&		--		&	$-$2.032\,$\pm$\,0.217	&	$-$3.629\,$\pm$\,0.179	&	1.137	&	111.9	&	--	&	--	&	16.3	\\
1002	&	S8UV049418	&	16464964-4552530	&	5940106281978531200	&	1.12	&	$-$0.004\,$\pm$\,0.135	&	$-$2.245\,$\pm$\,0.164	&	$-$3.991\,$\pm$\,0.134	&	1.040	&	58.8	&	16.3	&	18.8	&	14.8	\\		
1003	&	S8UV050318	&	16465234-4552032	&	5940106312026629376	&	0.99	&	0.314\,$\pm$\,0.099	&	$-$2.395\,$\pm$\,0.129	&	$-$3.688\,$\pm$\,0.108	&	1.129	&	27.7	&	17.0	&	19.8	&	15.6	\\		
1007	&	S8UV052780	&	--	&	5940106831734470272	&	1.02	&	0.306\,$\pm$\,0.088	&	$-$2.370\,$\pm$\,0.109	&	$-$3.572\,$\pm$\,0.093	&	1.126	&	32.3	&	16.1	&	19.7	&	14.6	\\		
1009	&	S8UV050785	&	--	&	5940106694295493376	&	1.04	&	0.077\,$\pm$\,0.087	&	$-$2.693\,$\pm$\,0.106	&	$-$3.801\,$\pm$\,0.092	&	1.124	&	23.3	&	16.2	&	19.2	&	14.6	\\		
1011	&	S8UV050317	&	--	&	5940106625576009344	&	1.38	&	0.533\,$\pm$\,0.124	&	$-$2.856\,$\pm$\,0.152	&	$-$3.811\,$\pm$\,0.127	&	1.145	&	40.3	&	16.8	&	19.4	&	15.3	\\		
1012	&	S8UV137923	&	--	&	5940106728646925952	&	1.09	&	0.221\,$\pm$\,0.109	&	$-$2.217\,$\pm$\,0.132	&	$-$3.933\,$\pm$\,0.113	&	1.109	&	18.1	&	17.0	&	20.7	&	15.4	\\		
1014	&	S8UV051272	&	--	&	5940106659927411456	&	1.07	&	0.204\,$\pm$\,0.094	&	$-$2.453\,$\pm$\,0.114	&	$-$3.623\,$\pm$\,0.099	&	1.160	&	33.9	&	17.0	&	19.9	&	15.5	\\		
1015	&	S8UV050784	&	16465798-4551408	&	5940106625576014976	&	0.90	&	0.201\,$\pm$\,0.090	&	$-$2.192\,$\pm$\,0.117	&	$-$3.740\,$\pm$\,0.094	&	1.084	&	33.5	&	16.2	&	19.5	&	14.7	\\		
1017	&	--	&	16465824-4550341	&	5940106827422726784	&	1.03	&	0.243\,$\pm$\,0.102	&	$-$2.454\,$\pm$\,0.124	&	$-$3.543\,$\pm$\,0.107	&	1.121	&	7.5	&	17.0	&	19.6	&	15.4	\\		
1018	&	--	&	16465824-4550570	&	5940106655624030720	&	1.00	&	0.334\,$\pm$\,0.094	&	$-$2.023\,$\pm$\,0.115	&	$-$3.883\,$\pm$\,0.099	&	1.113	&	26.5	&	16.6	&	20.0	&	15.0	\\		
1023	&	--	&	16470016-4551104	&	5940106659935760768	&	1.00	&	0.286\,$\pm$\,0.095	&	$-$2.243\,$\pm$\,0.115	&	$-$4.298\,$\pm$\,0.106	&	1.110	&	26.7	&	16.2	&	19.6	&	14.6	\\		
1024	&	S8UV051649	&	16470079-4551019	&	5940106763014978304	&	1.10	&	0.052\,$\pm$\,0.097	&	$-$2.542\,$\pm$\,0.116	&	$-$3.691\,$\pm$\,0.103	&	1.122	&	26.6	&	16.1	&	19.5	&	14.5	\\		
1025	&	--	&	16470075-4552048	&	5940106659935744512	&	1.05	&	$-$0.043\,$\pm$\,0.099	&	$-$2.379\,$\pm$\,0.122	&	$-$3.917\,$\pm$\,0.100	&	1.170	&	2.6	&	17.2	&	18.4	&	15.4	\\		
1026	&	2MIU38JC	&	16470103-4549490	&	5940106793069532544	&	1.05	&	0.181\,$\pm$\,0.106	&	$-$1.967\,$\pm$\,0.129	&	$-$3.360\,$\pm$\,0.111	&	1.120	&	12.9	&	17.1	&	20.4	&	15.6	\\		
1027	&	2MIU38K4	&	16470103-4550069	&	5940106797374730240	&	1.12	&	0.156\,$\pm$\,0.115	&	$-$2.272\,$\pm$\,0.140	&	$-$3.773\,$\pm$\,0.120	&	1.095	&	9.7	&	16.9	&	21.2	&	15.2	\\		
1032	&	S8UV052545	&	--	&	5940106797366425856	&	1.03	&	$-$0.020\,$\pm$\,0.127	&	$-$1.962\,$\pm$\,0.151	&	$-$3.903\,$\pm$\,0.130	&	1.112	&	12.1	&	17.5	&	21.0	&	16.0	\\		
1042	&	S8UV050292	&	--	&	5940105904013082240	&	1.15	&	0.846\,$\pm$\,0.358	&	$-$10.050\,$\pm$\,0.886	&	$-$12.600\,$\pm$\,0.488	&	1.454	&	276.1	&	18.7	&	--	&	--	\\		
1053	&	--	&	16470737-4548501	&	5940199912252322304	&	1.16	&	0.088\,$\pm$\,0.110	&	$-$2.282\,$\pm$\,0.138	&	$-$3.720\,$\pm$\,0.115	&	1.091	&	12.0	&	15.9	&	20.5	&	14.3	\\		
1056	&	S8UV051666	&	16470870-4551016	&	5940106041460464256	&	1.93	&	$-$0.546\,$\pm$\,0.267	&	$-$2.18\,$\pm$\,0.36	&	$-$1.69\,$\pm$\,0.31	&	0.992	&	17.7	&	16.5	&	20.3	&	14.8	\\		
1061	&	S8UV146498	&	16470974-4550402	&	5940199804872244864	&	0.97	&		--		&	$-$2.307\,$\pm$\,0.153	&	$-$4.283\,$\pm$\,0.129	&	1.091	&	164.4	&	--	&	--	&	16.0	\\
1064	&	S8UV138096	&	--	&	5940199843532840320	&	1.13	&	$-$0.054\,$\pm$\,0.152	&	$-$1.948\,$\pm$\,0.198	&	$-$3.972\,$\pm$\,0.159	&	1.050	&	6.7	&	17.5	&	21.8	&	15.8	\\		
1065	&	S8UV053673	&	16471160-4549226	&	5940199912252320128	&	0.98	&	0.146\,$\pm$\,0.095	&	$-$2.202\,$\pm$\,0.116	&	$-$4.099\,$\pm$\,0.101	&	1.085	&	44.9	&	15.3	&	19.1	&	13.7	\\		
1067	&	2MIU3H39	&	16471338-4549106	&	5940199946612058880	&	0.95	&	0.048\,$\pm$\,0.094	&	$-$1.993\,$\pm$\,0.121	&	$-$3.624\,$\pm$\,0.102	&	1.095	&	60.8	&	15.0	&	18.9	&	13.4	\\		
1069	&	S8UU115903	&	16472417-4553292	&	5940102742925551616	&	1.31	&	0.190\,$\pm$\,0.139	&	$-$2.319\,$\pm$\,0.186	&	$-$3.961\,$\pm$\,0.149	&	1.048	&	6.7	&	15.5	&	21.1	&	13.7	\\		
\hline
\end{tabular} \\
{\textbf{Notes:} The columns are as follows:
(1) Name of the source (the coordinates are listed in \citet{Clark19}); 
(2) designation of the source in the GSC 2.3 catalogue;
(3) designation of the source in the 2MASS PSC catalogue \citep{Skrutskie06};
(4) Identification of the Gaia EDR3 source;
(5) Renormalized Unit Weight Error;
(6) Parallax listed from the Gaia EDR3 and its error;
\citet{Lindegren21b};
(7)-(8) Proper motion in RA and Decl axis and their error;
(9) Pseudocolour;
(10) Minimum photometric error of the G, BP and RP-bands (in $\sigma$ units)
(11)-(13) Magnitude in the G, BP and RP-bands, respectively.
}
\end{table*}

\setlength{\tabcolsep}{6pt}
\end{landscape}
\end{onecolumn}
\setlength{\tabcolsep}{6pt}
\begin{table*}
\caption{List of known Westerlund 1 members with no astrometric information. This table is available online at the CDS.}
\label{table_Wd1members_noGaia}
\begin{tabular}{lccc|lccc}
\hline
\multicolumn{1}{c}{Name} & \multicolumn{1}{c}{GSC} &  \multicolumn{1}{c}{2MASS} &  \multicolumn{1}{c}{Gaia EDR3} & 
\multicolumn{1}{c}{Name} & \multicolumn{1}{c}{GSC} &  \multicolumn{1}{c}{2MASS} &  \multicolumn{1}{c}{Gaia EDR3} \\
\multicolumn{1}{c}{} &  \multicolumn{1}{c}{designation} &  \multicolumn{1}{c}{designation} &  \multicolumn{1}{c}{designation} & 
\multicolumn{1}{c}{} &  \multicolumn{1}{c}{designation} &  \multicolumn{1}{c}{designation} &  \multicolumn{1}{c}{designation} \\
\hline
W1	&	--	&	--	&	--	&	WR B	&	--	&	--	&	--	\\
W6b	&	--	&	--	&	--	&	WR C	&	--	&	--	&	--	\\
W8b	&	--	&	--	&	--	&	WR D	&	--	&	--	&	--	\\
W9	&	S8UV052695	&	16470414-4550312	&	5940106797374722688	&	WR G	&	--	&	--	&	--	\\
W14c	&	--	&	--	&	--	&	WR H	&	--	&	--	&	--	\\
W15	&	--	&	--	&	--	&	WR J	&	--	&	--	&	--	\\
W17	&	--	&	--	&	--	&	WR K	&	--	&	--	&	--	\\
W18	&	S8UV052696	&	16470570-4550506	&	--	&	WR O	&	2MIU39AQ	&	16470763-4552352	&	--	\\
W27	&	S8UV112081	&	--	&	5940106041460655744	&	WR Q	&	--	&	--	&	--	\\
W29	&	--	&	--	&	--	&	WR U	&	--	&	--	&	--	\\
W30	&	--	&	--	&	--	&	WR V	&	--	&	--	&	--	\\
W31	&	--	&	--	&	--	&	1001	&	S8UV137517	&	16464919-4553101	&	5940106178899314432	\\
W34	&	--	&	--	&	--	&	1004	&	--	&	--	&	--	\\
W36	&	S8UV111999	&	16470508-4550553	&	--	&	1008	&	--	&	--	&	--	\\
W37	&	--	&	--	&	--	&	1013	&	S8UV049740	&	--	&	--	\\
W38	&	--	&	--	&	--	&	1022	&	--	&	--	&	--	\\
W41	&	S8UV111985	&	--	&	5940106763014977152	&	1031	&	--	&	--	&	--	\\
W43a	&	--	&	--	&	--	&	1035	&	--	&	--	&	--	\\
W43b	&	--	&	--	&	--	&	1036	&	--	&	--	&	--	\\
W43c	&	S8UV111984	&	--	&	5940106007092379904	&	1039	&	--	&	--	&	--	\\
W44	&	--	&	--	&	--	&	1041	&	--	&	--	&	--	\\
W46a	&	--	&	--	&	--	&	1043	&	--	&	--	&	--	\\
W46b	&	--	&	--	&	--	&	1044	&	--	&	--	&	--	\\
W50b	&	S8UV052712	&	--	&	5940106763006683136	&	1045	&	S8UV050529	&	--	&	5940106007092303360	\\
W53	&	--	&	--	&	--	&	1047	&	S8UV049738	&	16470612-4552320	&	--	\\
W56b	&	--	&	--	&	--	&	1048	&	2MIU38UI	&	16470625-4551042	&	--	\\
W57c	&	--	&	--	&	--	&	1051	&	2MIU3H55	&	16470701-4549405	&	--	\\
W61a	&	S8UV051190	&	16470231-4551416	&	--	&	1054	&	--	&	--	&	--	\\
W61b	&	--	&	--	&	--	&	1055	&	--	&	--	&	--	\\
W70	&	S8UV052001	&	16470938-4550496	&	--	&	1057	&	--	&	--	&	--	\\
W72	&	--	&	--	&	--	&	1059	&	S8UV048805	&	16470907-4553204	&	--	\\
W75	&	2MIU38JS	&	16470892-4549585	&	--	&	1060	&	--	&	--	&	--	\\
W86	&	--	&	--	&	--	&	1066	&	--	&	--	&	--	\\
W237	&	S8UV050219	&	16470309-4552189	&	--	&	1068	&	S8UV050802	&	--	&	5940199603008974464	\\
W239	&	--	&	--	&	--	&		&		&		&		\\
\hline
\end{tabular} \\
{\textbf{Notes:} The columns are as follows:
(1) Name of the source; 
(2) designation of the source in the GSC 2.3 catalogue;
(3) designation of the source in the 2MASS PSC catalogue \citep{Skrutskie06};
(4) Identification of the source in the Gaia EDR3 catalogue (no astrometry is available).}
\end{table*}
\setlength{\tabcolsep}{6pt}

\bsp	
\label{lastpage}
\end{document}